\documentclass{jfm}

\usepackage{graphicx}
\usepackage{natbib}

\usepackage{amsmath}
\usepackage{subcaption}
\usepackage{bm}
\ifCUPmtlplainloaded \else
  \checkfont{eurm10}
  \iffontfound
    \IfFileExists{upmath.sty}
      {\typeout{^^JFound AMS Euler Roman fonts on the system,
                   using the 'upmath' package.^^J}%
       \usepackage{upmath}}
      {\typeout{^^JFound AMS Euler Roman fonts on the system, but you
                   dont seem to have the}%
       \typeout{'upmath' package installed. JFM.cls can take advantage
                 of these fonts,^^Jif you use 'upmath' package.^^J}%
      }
  \else
  \fi
\fi

\ifCUPmtlplainloaded \else
  \checkfont{msam10}
  \iffontfound
    \IfFileExists{amssymb.sty}
      {\typeout{^^JFound AMS Symbol fonts on the system, using the
                'amssymb' package.^^J}%
       \usepackage{amssymb}%
         \let\leq=\leqslant
         
      }{}
  \fi
\fi

\ifCUPmtlplainloaded \else
  \IfFileExists{amsbsy.sty}
    {\typeout{^^JFound the 'amsbsy' package on the system, using it.^^J}%
     \usepackage{amsbsy}}
    {}
\fi

\newsavebox{\astrutbox}
\sbox{\astrutbox}{\rule[-5pt]{0pt}{20pt}}

\usepackage{color}
\definecolor{light-gray}{gray}{0.5}
\definecolor{blue}{rgb}{0.0,0.0,1.0}
\definecolor{green}{rgb}{0.0,0.5,0.0}
\definecolor{red}{rgb}{1.0,0.0,0.0}
\definecolor{cyan}{rgb}{0.0,0.75,0.75}
\definecolor{magenta}{rgb}{0.75,0.0,0.75}
\definecolor{yellow}{rgb}{0.75,0.75,0.0}

\newcommand{\avg}[1]{\langle{#1}\rangle}
\newcommand{\sdot}{\cdot}

\newcommand{\grad}{\bm \nabla}
\newcommand{\pd}{\partial}

\title[Forcing-dependent dynamics and emergence of helicity in rotating turbulence]{Forcing-dependent dynamics and emergence of helicity in rotating turbulence}
\author[V. Dallas and S. M. Tobias]%
{Vassilios Dallas%
  \thanks{Email address for correspondence: v.dallas@leeds.ac.uk}\ns
and Steven M. Tobias}

\affiliation{Department of Applied Mathematics, University of Leeds, Leeds LS2 9JT, UK}

\pubyear{2016}
\volume{650}
\pagerange{119--126}
\date{?; revised ?; accepted ?. - To be entered by editorial office}
\begin{document}

\maketitle

%abstract of no more than 250 words, which provides a summary of the main aims and results. 
\begin{abstract}
The effects of large scale mechanical forcing on the dynamics of rotating turbulent flows are studied by means of numerical simulations, varying systematically the nature of the mechanical force in time. We demonstrate that the statistically stationary solutions of these flows depend on the nature of the forcing mechanism. 
%For strong enough rotation these flows with a forcing that adds a preferred direction to the propagation of the inertial waves bifurcate from a non-helical state to a helical state even though the forcing is non-helical. 
Rapidly enough rotating flows with a forcing that has a persistent direction relatively to the axis of rotation bifurcate from a non-helical state to a helical state despite the fact that the forcing is non-helical. We find that the nature of the mechanical force in time and the emergence of helicity have direct implications on the cascade dynamics of these flows, determining the anisotropy in the flow, the energy condensation at large scales and the power-law energy spectra that are consistent with previous findings and phenomenologies under strong and weak-wave turbulent conditions.
\end{abstract}


\section{\label{sec:intro}Introduction}
The effects of the Coriolis force on a turbulent fluid flow become important at sufficiently high rotation rates altering its dynamics \citep{tritton88}. Experiments and simulations reveal that fast rotation renders the flow quasi-two-dimensional (quasi-2D), since fast rotation suppresses the velocity gradients along the axis of rotation as shown by the Taylor-Proudman theorem \citep{proudman16,taylor17}. %for flows with a Coriolis force much greater than the inertial force. 
Under such conditions, the flow sustains inertial waves whose frequency is proportional to the rotation rate \citep{lighthill65,greenspan68}. %It is also observed that rotation introduces a transfer of energy from the forcing scale to the largest scale of the system (inverse energy cascade) and the inertial waves alter the energy distribution among scales by inhibiting the transfer of energy to the small scales (forward energy cascade) \citep{sagautcambon08,davidson13}.

The interplay between inertial waves and eddies in rotating fluids makes the problem of rotating turbulence very rich. 
%For example, different results are considered for different values of the involved control parameters of the problem. This has led to the commencement of many experimental studies at a wide range of parameters
Many experimental studies at a wide range of parameters have elucidated the dynamics of such flows \citep{hopfingervanheijst93,ruppertetal05,davidsonetal06,bewleyetal07,vanbokhovenetal09,moisyetal11,yarometal13,campagneetal14}. Many numerical studies have also been carried out on rotating turbulence. The regime for which both the turbulence is fully developed (large Reynolds numbers) and the flow is fast rotating (small Rossby numbers) puts strong restrictions on the scale separation requirements in 
simulations. Therefore, most of the early numerical investigations were focused on decaying rotating turbulence \citep{bartelloetal94,hossain94,cambonetal97,morinishietal01,teitelbaummininni11,yoshimatsuetal11}.
More recent studies have been performed on forced rotating turbulence both at large \citep{yeungzhou98,mininnipouquet10,mininnietal12} and small scales; the latter to study the dynamics of the inverse cascade \citep{smithetal96,pouquetetal13,deusebioetal14}. The computational costs prevent in general the exhaustive coverage of the parameters space with most numerical studies reaching either large Reynolds numbers and moderate Rossby numbers or moderate Reynolds numbers and small Rossby numbers. 
%Note that these simulations have not reached a steady state because very long integration times are required. %With 
%The exception is a very recent study %with a 184 numerical simulations 
%which reached steady states and covered extensively a fairly large portion of the parameters space \citep{alexakis15}.
Note that these simulations, with the exception of a very recent study which reached steady states and covered extensively a fairly large portion of the parameters space \citep{alexakis15}, have not reached statistically stationary solutions %a steady state 
because very long integration times are required.

Although these studies were numerous and an adequate portion of the parameters space was covered there are still disparate results in different cases --- for example, different power law spectra with $E(k) \propto k^{-5/3}$, $k^{-2}$ and $k^{-5/2}$, supported theoretically by strong and weak-wave turbulence phenomenologies \citep{k41a,zhou95,galtier03,pouquetmininni10}. Moreover, a recent investigation showed sensitivity to forcing in that different results for the large scales of a flow that was forced at intermediate scales \citep{senetal12} when energy was injected exclusively to the quasi-2D component of the flow and when it was injected solely to the inertial waves. The injection of helicity into the flow has also shown alterations on the behaviour of the cascade \citep{pouquetmininni10}. Forcing dependent dynamics have been observed in various other systems such as in two-dimensional (2D) turbulence \citep{braccowilliams10,boffettaecke12}, in beta-plane turbulence \citep{maltrudvallis91} and in magnetohydrodynamic turbulence \citep{da15}.

The present work focuses on the effects of the mechanical force on the dynamics of rotating flows by means of numerical simulations, varying systematically the memory time scale of the mechanical force (i.e. the time scale of which the phases of the force are randomised).
%the time scale of randomising the phases of the external force. 
The behaviour of different mechanical forcing mechanisms on the flows is also considered for different rotation rates. To the best of the authors' knowledge this is the first study of forced rotating flows in the steady state regime where the effects of a large scale external force on the dynamics are studied extensively.

The paper is structured as follows. All the necessary details of the formulation of our direct numerical simulations (DNS) of forced rotating turbulence are provided in Sec. \ref{sec:dns}. Section \ref{sec:taum} analyses the dynamics of the flows with different memory time scales of the forcing mechanism for a given Rossby number. Here, we also focus on the spontaneous emergence of helicity in our flows and its influence on the anisotropy and on the spectral dynamics. In Sec. \ref{sec:Rossby} we describe the Rossby number dependence on flows with different type of forcing mechanisms %\newer{A dispersion relation taking into account the effect of the mechanical forcing is derived also in Sec. \ref{sec:Rossby}} 
and we justify the spontaneous mirror-symmetry breaking in our flows even though net helicity is not injected directly. Finally, in Sec. \ref{sec:end}, we conclude by summarising our findings and we discuss the implications of our work.

\section{\label{sec:dns}Numerical Simulations}
In this study, we consider the three-dimensional (3D) incompressible Navier-Stokes equations in a rotating frame of reference
\begin{equation}
 \pd_t \bm u + \bm \omega \times \bm u + 2\bm\Omega \times \bm u = - \grad P + \nu \grad^2 \bm u + \bm f,
 \label{eq:ns}
\end{equation}
where $\bm u$ is the velocity field, $\bm \omega = \grad \times \bm u$ is the vorticity, $P$ is the pressure, $\nu$ is the kinematic viscosity, and $\bm f$ is an external mechanical force. In a Cartesian domain, we choose the rotation axis to be in the $z$ direction with $\bm \Omega = \Omega \bm e_z$, where $\Omega$ is the rotation frequency. In the ideal case of $\nu = 0$ and $\bm f = 0$, Eq. \eqref{eq:ns} conserves the energy $E = \frac{1}{2}\avg{|\bm u|^2}$ (where $|\sdot|$ stands for the $L_2$-norm) and the helicity $H = \avg{\bm u \sdot \bm \omega}$ with the angular brackets denoting a spatial average unless indicated otherwise. 

The external mechanical forcing in Eq. \eqref{eq:ns} is given by
 \begin{equation}
 \bm f = f_0 \sum_{k_f}
   \begin{pmatrix}
   \sin(k_f y + \phi_y) + \sin(k_f z + \phi_z) \\
   \sin(k_f x + \phi_x) + \sin(k_f z + \phi_z) \\
   \sin(k_f x + \phi_x) + \sin(k_f y + \phi_y) \\
 \end{pmatrix}.
  \label{eq:forcing}
 \end{equation}
For all the runs the forcing amplitude is normalised such that $\bm f / |\bm f| = f_0 = 1$ and the phases $\phi_x$, $\phi_y$, $\phi_z$ are randomised for the forcing wavenumbers $k_f = 2$ to 4 every $\tau_m$, the so called memory time scale, which is one of the contol parameters in our study. In the limit of $\tau_m \rightarrow 0$ we have essentially a random delta correlated in time forcing with the phases randomised at each time step, whereas when we choose $\tau_m = \infty$ we randomise the phases only at $t=0$ in the duration of the runs and hence we apply a time-independent forcing. Note that our forcing mechanism 
%chosen such that zero net helicity $\avg{\bm f \sdot \grad \times \bm f} = 0$ is injected into the flow.
has $\bm f \sdot \grad \times \bm f \neq 0$ pointwise in space but it is non-helical on average $\avg{\bm f \sdot \grad \times \bm f} = 0$.

Now, if we write the wavenumbers in the 3D Fourier space using cylindrical coordinates, we have $\bm k = (\bm k_\perp, \bm k_\parallel)$, with $k_\perp = \sqrt{k_x^2 + k_y^2}$ and $k_\parallel = |k_z|$. Then, the 2D modes %in the sense that they are 
(i.e. independent of $z$) in Fourier space can be denoted as $\bm u(\bm k_\perp)$ and the 3D or wave modes as $\bm u(\bm k)$. Then, in this setting the time-independent forcing ($\tau_m = \infty$) excites two 2D modes in the $k_x$ and $k_y$ axis of the Fourier space and one 3D mode in the $k_z$ axis in the Fourier shell of amplitude $|\bm k|=k_f$. The random-in-time forcing ($\tau_m \rightarrow 0$) excites also the same modes but since the phases are random, this mechanism is essentially isotropic in contrast to the time-independent forcing.

The relevant dimensionless control parameters of our problem are defined based on the forcing amplitude. So, the Reynolds number is given by $Re_f = U/(k_{min}\nu)$ and the Rossby number by $Ro_f = Uk_{min}/(2\Omega)$ where $U = (f_0/k_{min})^{1/2}$. Using these definitions $Re_f^2$ is essentially the Grashof number and $Ro_f$ the ratio of the rotation period $\tau_w \propto \Omega^{-1}$ to the eddy turnover time $\tau_{NL} = (Uk_{min})^{-1}$. Note that $Re_f$ and $Ro_f$ are control parameters that they do not require knowledge of the solution to be evaluated and are useful for comparison with body-forced numerical simulations or experiments. %The large scale turnover time %associated with the forcing amplitude 
%is defined as %$\tau_f = (f_0k_{min})^{-1/2}$. 
%$\tau_{NL} = (Uk_{min})^{-1}$. 
All the control parameters of our DNS are listed in Table \ref{tbl:dnsparam}.

%%%%%%%%%%%%%%%%%%%%%%%%%%%%%%%%%%%%%%%%%%%%%%%%%%%%%%%%%%%%%%%%%%%%%%%%%%%%%%%%%%%%%%%%%%%%%%%%%%%

\begin{table}
  \begin{center}
\def~{\hphantom{0}}
    \begin{tabular}{l*{16}{c}}
     $\tau_m/\tau_{NL}$ & 0.5  & 0.5  & 0.5 & 0.5 & 0.5 & 0.5 & 4.0 & 32.0 & 128.0 & $\infty$ & $\infty$ & $\infty$ & $\infty$ & $\infty$ & $\infty$ & $\infty$ \\
     $Ro_f$                         & 0.01 & 0.05 & 0.1 & 0.1 & 0.2 & 0.5 & 0.1 & 0.1  & 0.1   & 0.01     & 0.05     & 0.1      & 0.1      & 0.2      & 0.33     & 0.5      \\
     $Re_f$                         & 333  & 333  & 333 & 714 & 333 & 333 & 333 & 333  & 333   & 333      & 333      & 333      & 714      & 333      & 333      & 333      \\
     $\Omega$           & 50.0 & 10.0 & 5.0 & 5.0 & 2.5 & 1.0 & 5.0 & 5.0  & 5.0   & 50.0     & 10.0     & 5.0      & 5.0      & 1.0      & 1.5      & 1.0      \\
     $\nu \; (\times 10^{-3})$             & $3.0 $ & $3.0 $ & $3.0 $ & $1.4 $ & $3.0 $ & $3.0 $ & $3.0 $ & $3.0 $ & $3.0 $ & $3.0 $ & $3.0 $ & $3.0 $ & $1.4 $ & $3.0 $ & $3.0 $ & $3.0 $ \\
     $N$                & 256  & 256  & 256 & 512 & 256 & 256 & 256 & 256  & 256   & 256      & 256      & 256      & 512      & 256      & 256      & 256 \\
   \end{tabular}
  \caption{Numerical control parameters of the DNS.} %Here we present the memory time scale in a dimensionless form by normalising with the %large scale %forcing time scale 
%turnover time $\tau_{NL}$.}
  \label{tbl:dnsparam}
   \end{center}
\end{table}

%%%%%%%%%%%%%%%%%%%%%%%%%%%%%%%%%%%%%%%%%%%%%%%%%%%%%%%%%%%%%%%%%%%%%%%%%%%%%%%%%%%%%%%%%%%%%%%%%%%

Using the pseudo-spectral method, we numerically integrate Eq. \eqref{eq:ns} in a periodic box of size $2\pi$, satisfying the incompressibility condition $\grad \sdot \bm u = 0$. The time derivatives are estimated using a third-order Runge-Kutta scheme. Aliasing errors are removed using the $2/3$ dealiasing rule and as a result the minimum and maximum wavenumbers are $k_{min}=1$ and $k_{max}=N/3$, respectively, where $N$ is the number of grid points in each Cartesian coordinate. For more details on the numerical code see \citep{mpicode05}.

%
% \section{\label{sec:res}Memory time scale dependence}
% \section{\label{sec:taum}Dependence on the memory time scale}
\section{\label{sec:taum}Forcing-dependent dynamics}
\subsection{\label{sec:time}Time evolution}
\begin{figure}
 \centering
 \begin{subfigure}{0.49\textwidth}
   \includegraphics[width=\textwidth]{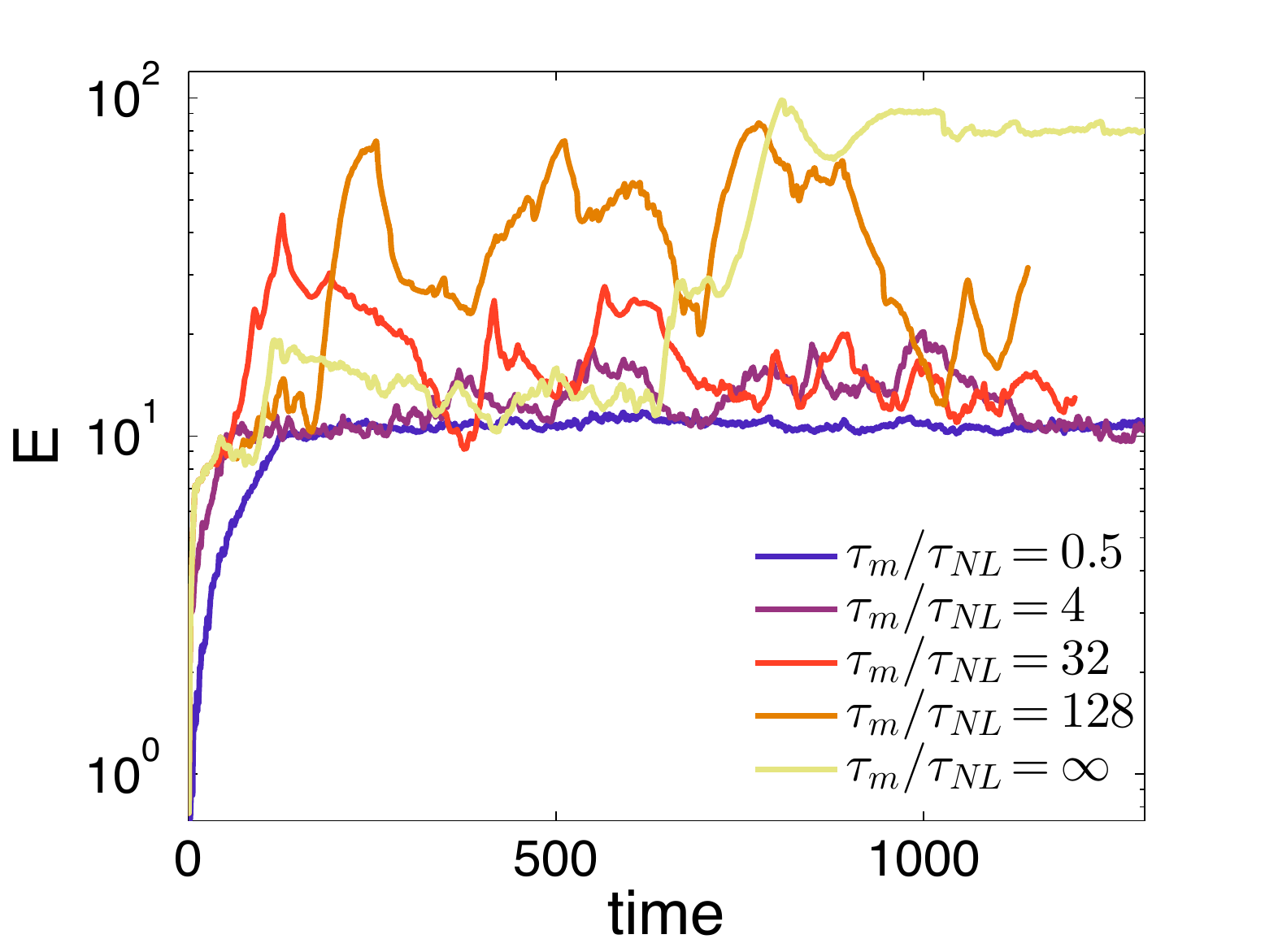}
   \caption{}
   \label{fig:energy}
 \end{subfigure}
 \begin{subfigure}{0.49\textwidth}
   \includegraphics[width=\textwidth]{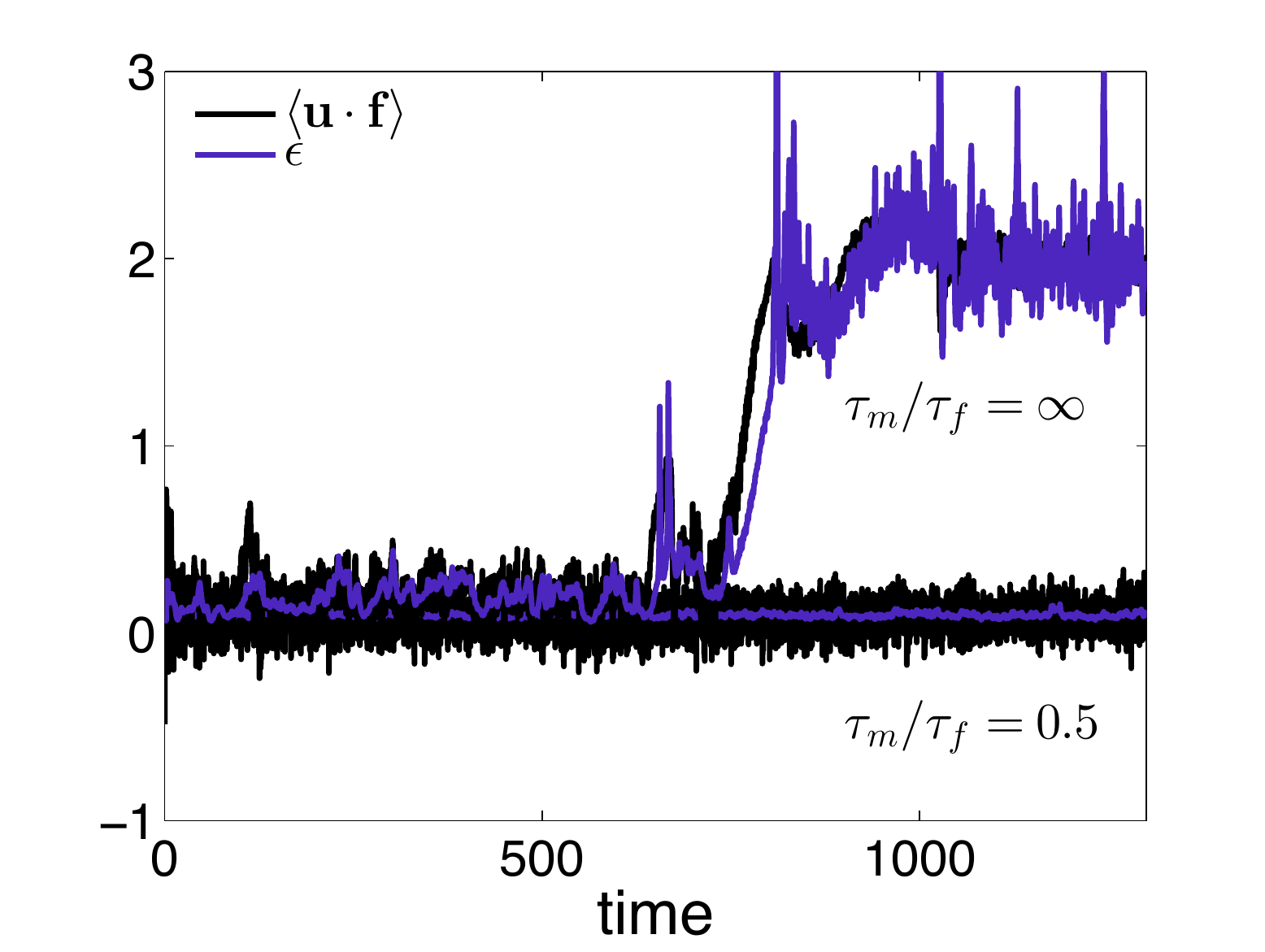}
   \caption{}
   \label{fig:injection}
 \end{subfigure} 
  \caption{(Color online) Time series of (a) the energy $E$ for the flows with different memory time scales $\tau_m$ of the forcing and of (b) the rates of energy dissipation $\epsilon$ and injection $\avg{\bm u \sdot \bm f}$ for the flows with $\tau_m/\tau_{NL} = 0.5$ and $\infty$ at Rossby number $Ro_f = 0.1$ and Reynolds number $Re_f = 333$}
%  \label{fig:energy}
  \label{fig:timeseries}
 \end{figure}

Figure \ref{fig:energy} shows the temporal evolution of the energy $E$ for flows with $Ro_f = 0.1$ and different memory time scale $\tau_m$ of the forcing.
As $\tau_m$ increases, we observe a gradual increase of the amplitude of the energy up to an order of magnitude. The time-series for $\tau_m/\tau_{NL} > 4$ are characterised by large signal variations, which %force us inevitably to carry out 
require extremely long time-integrations 
%striving for converged statistics. For this reason, we are restricted 
restricting our runs to moderate Reynolds numbers. %size numerical resolutions. 
Note that even for low $\tau_m$ a steady state is reached after a transient that lasts for about 50 to 100$\tau_{NL}$ turnover times indicating how expensive computationally is to reach a steady state regime in rotating flows forced at large scales.

The temporal evolution of the energy dissipation rate $\epsilon = \nu\avg{|\bm \omega|^2}$ and the energy injection rate $\avg{\bm u \sdot \bm f}$ are presented in Fig. \ref{fig:injection}. For clear illustration purposes we choose to plot only the two extreme cases of the flows with the highly random-in-time forcing ($\tau_m/\tau_{NL} = 0.5$) and the time-independent forcing ($\tau_m/\tau_{NL} = \infty$) at $Ro_f = 0.1$.
% \begin{figure}
% \centering
%  \includegraphics[width=0.5\textwidth]{Plots/uf-eps_vs_time.eps}
%  \caption{(Color online) Time series of the rate of energy dissipation $\epsilon$ and the injection $\avg{\bm u \sdot \bm f}$ for the flows with $\tau_m/\tau_{NL} = 0.5$ and $\infty$ at Rossby number $Ro_f = 0.1$ and Reynolds number $Re_f = 333$}
%  \label{fig:injection}
% \end{figure}
As we saw at relatively early times the two flows reach a steady state (see Fig. \ref{fig:energy}) and therefore the balance $\epsilon = \avg{\bm u \sdot \bm f}$ is satisfied with both flows having %more or less 
the same rates of energy injection and dissipation. However, after a very long time period ($\sim$ 600$\tau_{NL}$ turnover times) the flow with the time-independent forcing deviates to %a new transient regime to reach 
a new statistically steady state. This happens when $\bm u$ becomes correlated with $\bm f$ and then the flow adjusts its dissipation rate such that a new steady state is achieved. 

This adjustment of the dissipation rate by the flow owing to the increase of the correlation between the external mechanical force and the velocity field explains why the energy increases as we increase the memory time scale of the forcing. This is a very interesting property of rotating flows from a practical point of view if one wants to minimise or maximise the energy dissipation rate in a potential application such as in turbomachinery. %perhaps one would like to see if this is also true in turbulent convection.

\subsection{\label{sec:helicity}The role of helicity}
\begin{figure}
 \centering
 \begin{subfigure}{0.49\textwidth}
   \includegraphics[width=\textwidth]{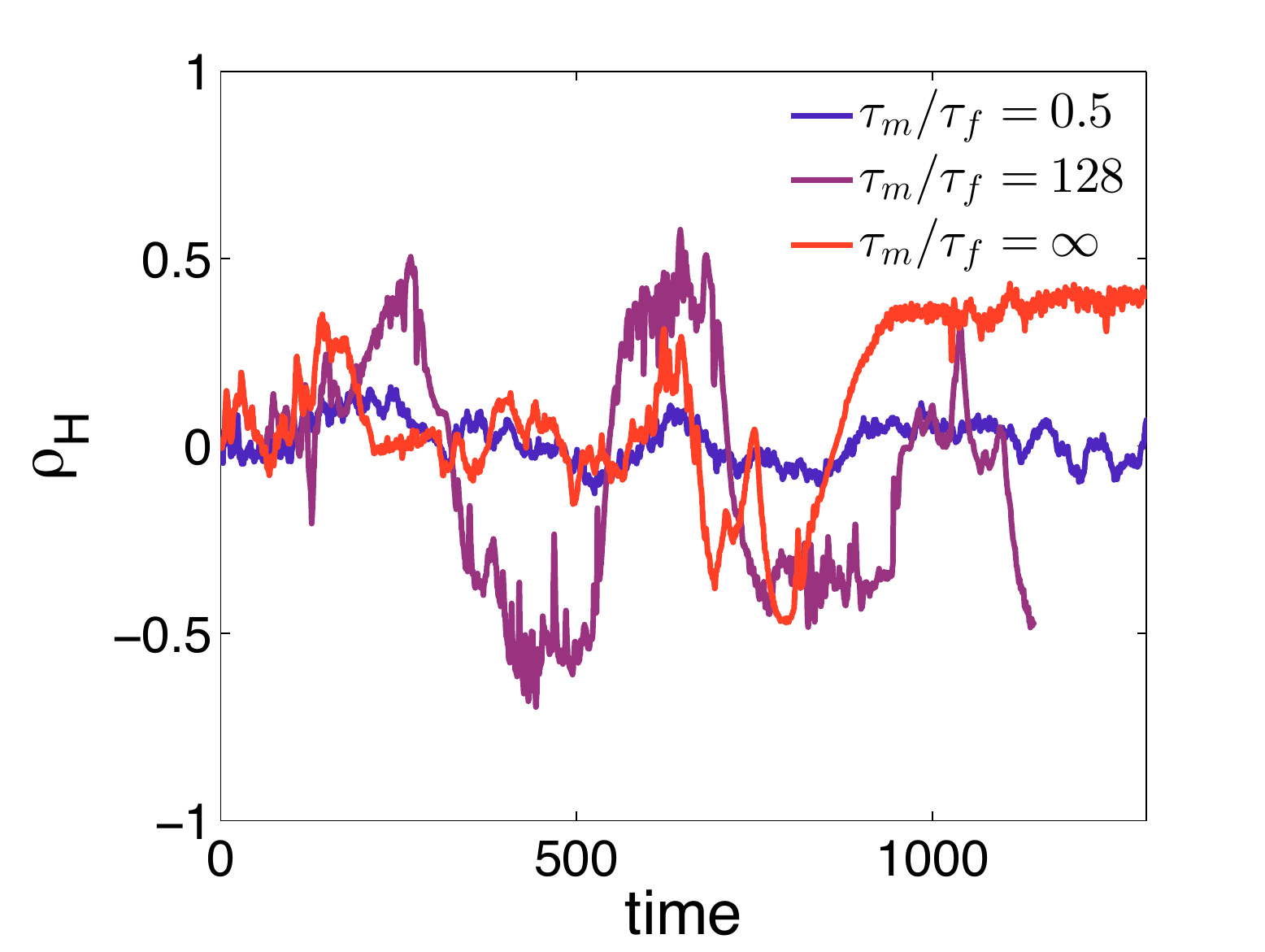}
   \caption{}
   \label{fig:hela}
 \end{subfigure}
 \begin{subfigure}{0.49\textwidth}
   \includegraphics[width=\textwidth]{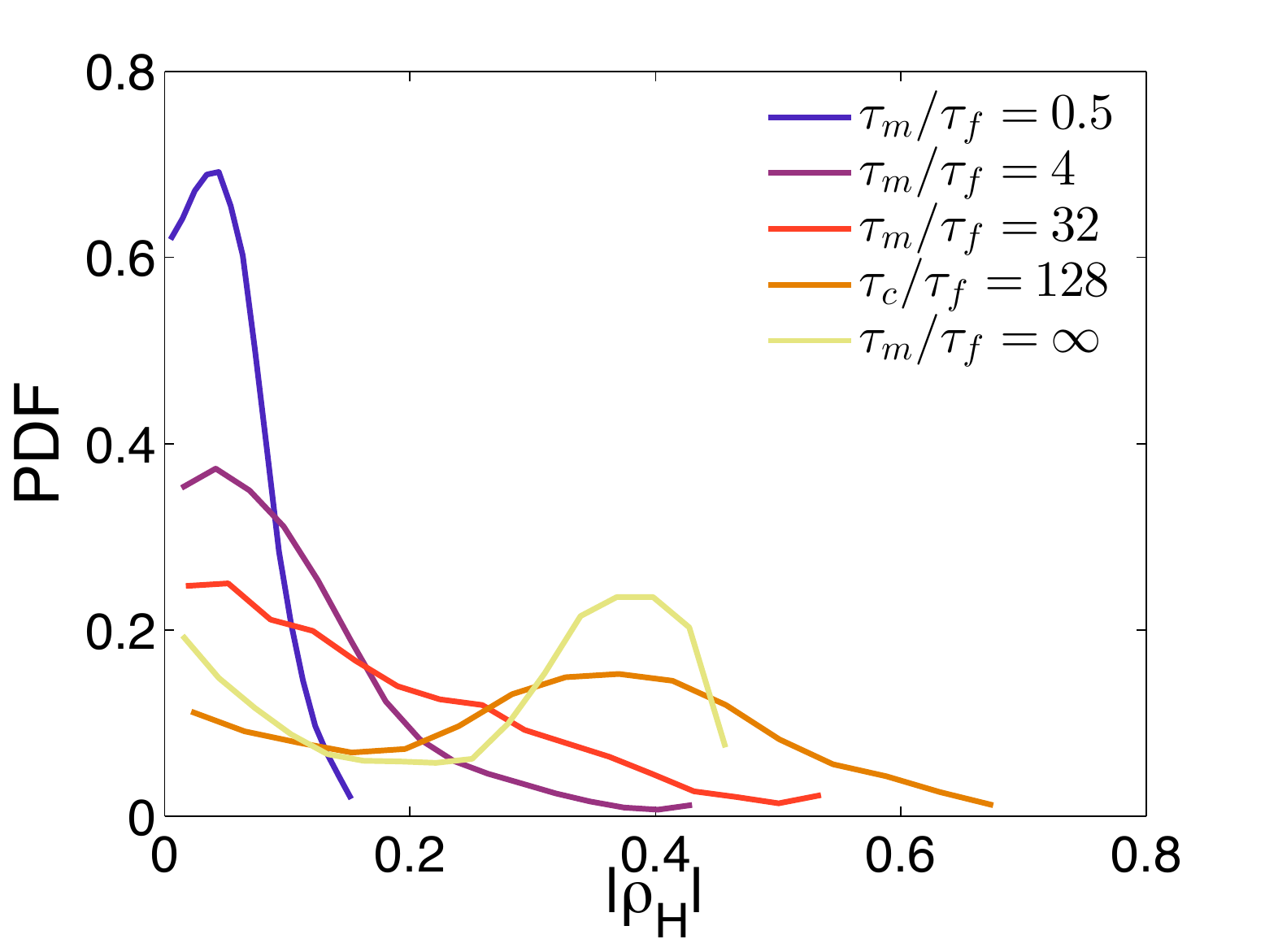}
   \caption{}
   \label{fig:helb}
 \end{subfigure}
  \caption{(Color online) (a) Time series of relative helicity $\rho_H$ and (b) Probability Density Function (PDF) of the absolute value of $\rho_H$ for the flows with different forcing memory time scale at Rossby number $Ro_f = 0.1$ and Reynolds number $Re_f = 333$}
  \label{fig:helicity}
 \end{figure}

Helicity is common in real flows and it can be created, for example, in planetary atmospheres in the presence of rotation and stratification 
\citep{moffatt78,tobias09,marinoetal13}. In homogenenous non-rotating turbulence it is expected that the helicity spectrum cannot develop if it is initially zero \citep{andrelesieur77} or if an external mechanism does not inject net helicity \citep{dfa15}. In our runs zero net helicity is injected into the flow. %$\avg{\bm f \sdot \grad \times \bm f} = 0$. 
Nevertheless, for a given rotation rate (i.e. $Ro_f = 0.1$) we observe that the relative helicity $\rho_H = H/(|\bm u| |\bm \omega|)$ increases as the memory time scale of the forcing increases (see Fig. \ref{fig:helicity}). 
%Note, however, that even though the injection rate of net helicity is zero, the forcing injects helicity pointwise in space (i.e. $\bm f \sdot \grad \times \bm f \neq 0$ at each point in space).

It is apparent from Fig. \ref{fig:helicity} that the mirror-symmetry breaking depends on the value of $\tau_m$. Figure \ref{fig:hela} shows
$\rho_H$ to be almost zero at early times for all the flows and as $\tau_m$ increases the mirror-symmetry breaks at later times only for long enough $\tau_m$. %having to integrate for very long time scales for the $\tau_m/\tau_{NL} = \infty$ case in order to see a significant amount of relative helicity.
We observe that helicity %is created 
emerges in the flow as soon as $\tau_m$ becomes of the order of the eddy turnover time, %= (k_{min}U)^{-1}$ 
i.e. $\tau_m/\tau_{NL} \sim \mathcal O(1)$. To analyse further this behaviour of $\rho_H$ we plot the Probability Density Function (PDF) of the time series of the absolute value of relative helicity in Fig. \ref{fig:helb}. This plot shows an increase in the mean value of $|\rho_H|$ and also a broadening of the tails of the PDFs for longer memory time scales. So, unlike in homogeneous turbulence, helicity can be created in rotating flows by an external force with a sufficiently long memory time scale, even though net helicity is not injected directly into the flow.

To determine whether the breaking of mirror-symmetry, that distinguishes flows with highly random-in-time and time-independent forcings, remains at higher Reynolds numbers we carried out simulations for the two extreme cases of $\tau_m/\tau_{NL} = 0.5$ and $\infty$ at $Re_f = 714$ and $Ro_f = 0.1$.
Our numerical simulations confirm the persistence of this behaviour at higher Reynolds numbers. We therefore analyse these higher Reynolds number runs in order to gain futher insight on the effects of helicity on the flow.

\begin{figure}
 \centering
% \begin{subfigure}{0.23\textwidth}
 \begin{subfigure}{0.35\textwidth}
   \includegraphics[width=\textwidth]{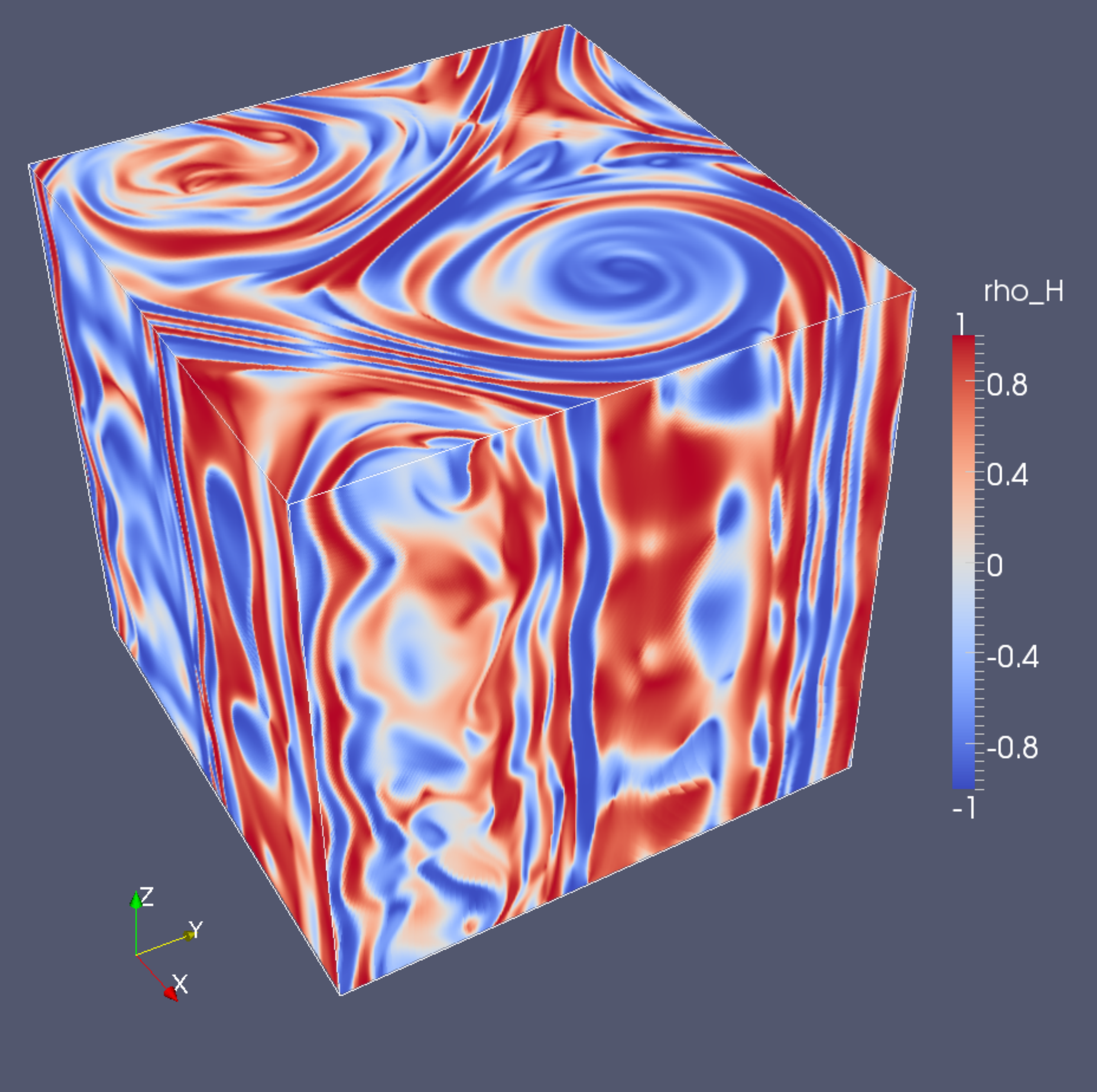}
   \caption{}
   \label{fig:vizhela}
 \end{subfigure}
 \hspace{1.0cm}
% \begin{subfigure}{0.23\textwidth}
 \begin{subfigure}{0.35\textwidth}
   \includegraphics[width=\textwidth]{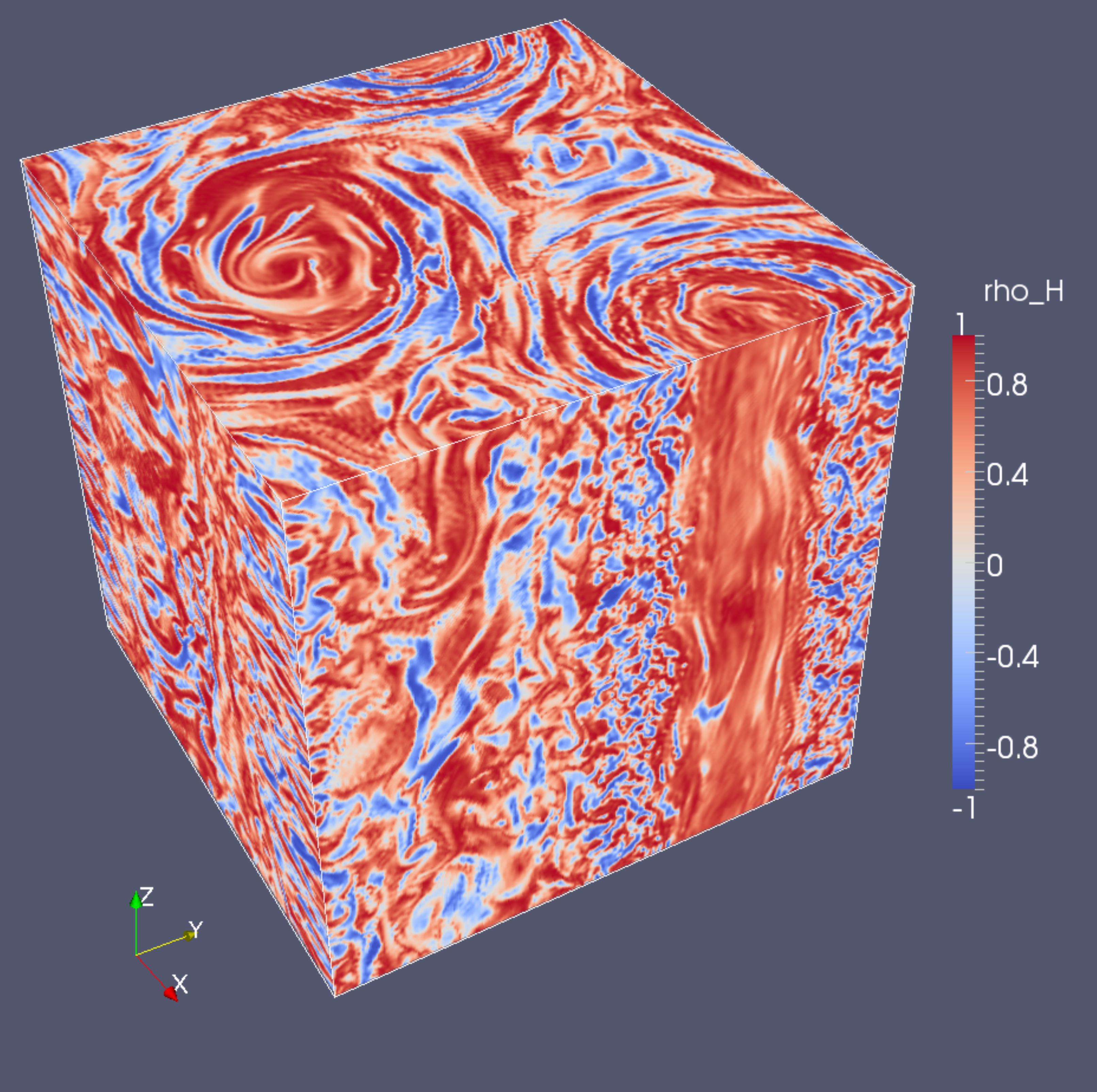}
   \caption{}
   \label{fig:vizhelb}
 \end{subfigure}
  \caption{(Color online) Visualisations of the relative helicity $\rho_H$ for the flows at Rossby number $Ro_f = 0.1$ and Reynolds number $Re_f = 714$ with (a) $\tau_m/\tau_{NL} = 0.5$, $\rho_H = -0.018$ and (b) $\tau_m/\tau_{NL} = \infty$, $\rho_H = 0.45$}
  \label{fig:visualisations}
 \end{figure}

Visualisations of the relative helicity of the %these high Reynolds number 
flows with $Re_f = 714$ are presented in Figs. \ref{fig:vizhela} and \ref{fig:vizhelb} with the %space average 
value of $\rho_H = -0.018$ and 0.45 for $\tau_m/\tau_{NL} = 0.5$ and $\infty$, respectively. %Note that Fig. \ref{fig:vizhelb} was chosen to show a time moment that $\rho_H$ reaches its maximum value in the duration of the run. 
%The red colour indicates positive helicity, while the blue colour indicates negative helicity.
The red and blue colour in Fig. \ref{fig:visualisations} indicate right-hand (positive helicity) and left-hand (negative helicity) circularly polarized helical waves, respectively.
% Visualisations
An instructive way to explain this further is to decompose the velocity field into circularly polarised helical waves \citep{constantinmajda88,waleffe92}
\begin{equation}
\bm u(\bm x,t) = \bm h_{\pm}(\bm k)e^{i(\bm k \sdot \bm x - \omega_{\pm}t)},  
\end{equation}
where $i\bm k$, $\bm h_+$ and $\bm h_-$ are the linearly independent eigenvectors of the curl operator, i.e. $i\bm k \times \bm h_{\pm} = \pm |\bm k| \bm h_{\pm}$. 
%, where $\bm h_{\pm}(\bm k) = \hat p \times \hat k \pm i \hat p$ with $\bm p(\bm k) = \hat z \times \bm k$. 
These complex eigenvectors are orthogonal to each other and are fully helical. So, now $\widehat{\bm u}(\bm k)$ can be expressed as a linear combination of the eigenvectors $\bm h_+$ and $\bm h_-$ only as follows 
\begin{equation}
\widehat{\bm u}(\bm k,t) = u_+(\bm k,t)\bm h_+(\bm k) + u_-(\bm k,t)\bm h_-(\bm k)
\end{equation}
since $\bm k \sdot \widehat{\bm u}(\bm k) = 0$.
Then, the helicity can be separated into modes of positive and negative helicity, viz. 
\begin{align}
H &= \sum_{\bm k} \widehat{\bm u}(\bm k) \sdot \widehat{\bm \omega}^*(\bm k) \nonumber \\
  &= \sum_{\bm k} k(|u_+(\bm k)|^2-|u_-(\bm k)|^2) \nonumber \\ 
  &= k(E_+ - E_-) = H_+ - H_-,
\end{align}
where $^*$ denotes the complex-conjugate. %So, the red and blue colour in Fig. \ref{fig:visualisations} can be seen as right-hand and left-hand circularly polarized helical waves, respectively.

The nature of the forcing is clearly imprinted on the flow structure in Fig. \ref{fig:visualisations}. The flow with the highly random-in-time forcing (see Fig. \ref{fig:vizhela}) gives a quasi-2D flow with two large columnar vortices, typical at low Rossby numbers due to the Taylor-Proudman theorem. These two vortices are governed by helical waves of opposite polarity. %In other words, $H_{\pm} = kE_{\pm}$ refer to the eigenfunctions of the curl operator, corresponding to left-hand (-) and right-hand (+) circularly polarised helical waves.
On the other hand, the flow with the time-independent forcing is characterised by helical waves of opposite polarity that travel within the flow breaking the %quasi-two-dimensionality 
quasi-2D behaviour at small and intermediate scales that is imposed by rotation (see Fig. \ref{fig:vizhelb}). Note that the two large scale vortices are still present but this time they have the same sign of helicity on average. %In this case the flow is expexted to be less anisotropic. 
Similar effects of helicity have also been observed on an elder study of decaying rotating turbulence \citep{morinishietal01}.

\begin{figure}
 \centering
% \begin{subfigure}{0.23\textwidth}
  \begin{subfigure}{0.35\textwidth}
   \includegraphics[width=\textwidth]{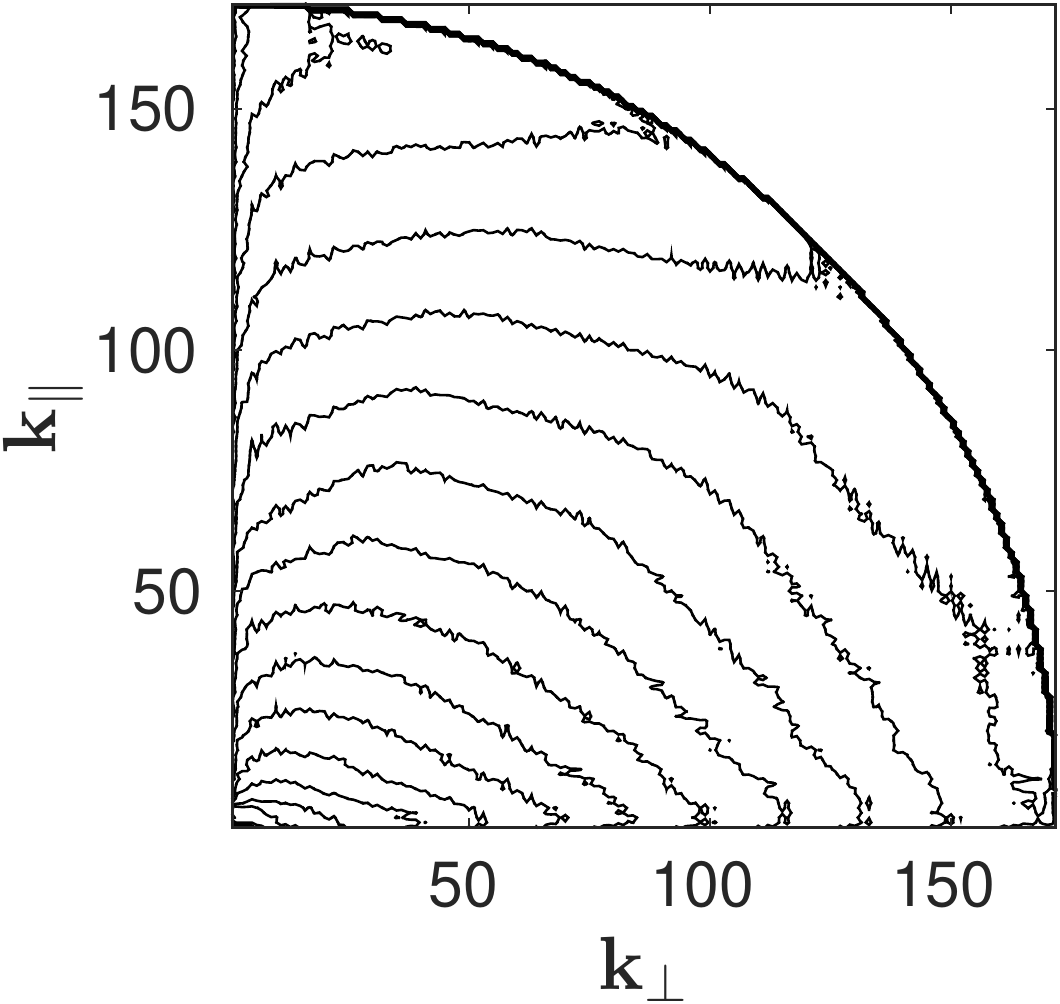}
   \caption{}
   \label{fig:2Dspeca}
 \end{subfigure} 
 \hspace{1.0cm}
% \begin{subfigure}{0.23\textwidth}
 \begin{subfigure}{0.35\textwidth}
   \includegraphics[width=\textwidth]{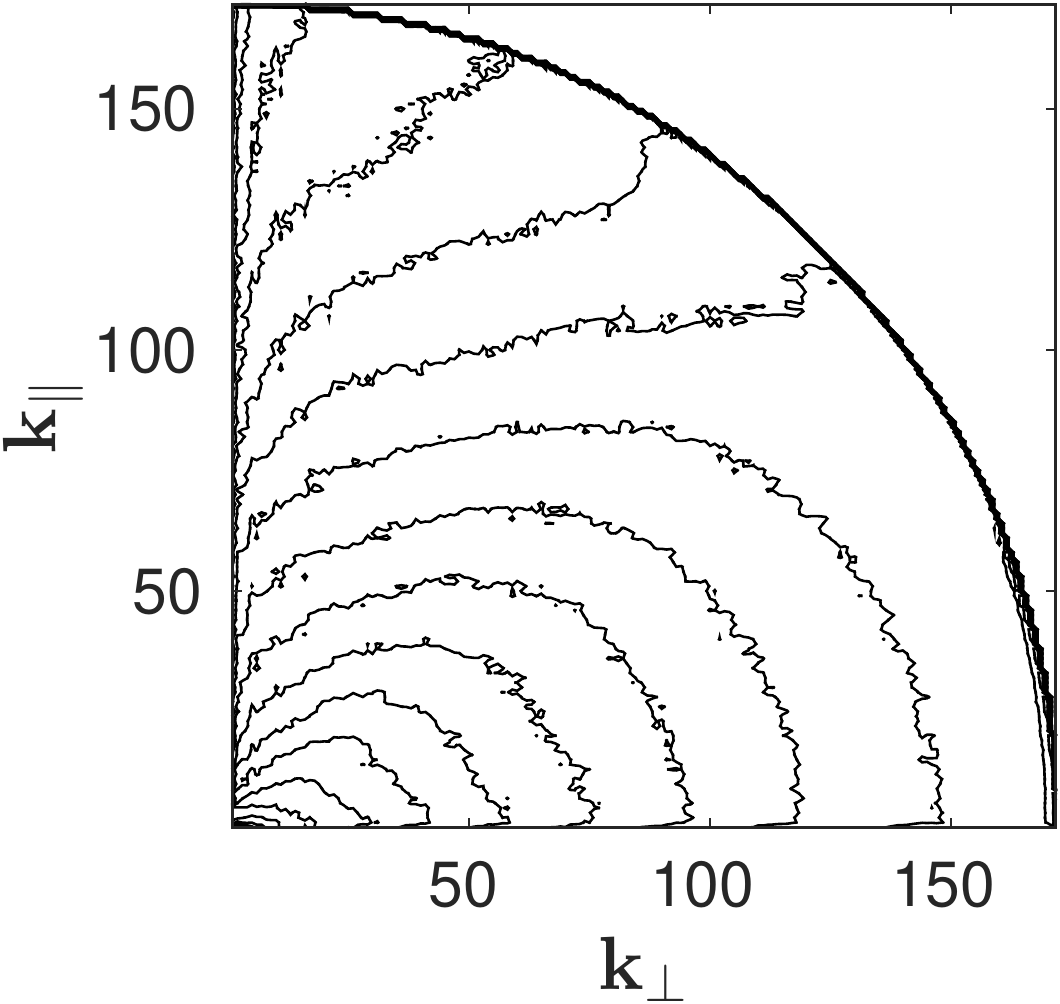}
   \caption{}
   \label{fig:2Dspecb}
 \end{subfigure}
  \caption{(Color online) The two-dimensional energy spectrum $E_{2D}(k_\perp,k_\parallel)$ for the flows at Rossby number $Ro_f = 0.1$ and Reynolds number $Re_f = 714$ with (a) $\tau_m/\tau_{NL} = 0.5$, $\rho_H = -0.018$ and (b) $\tau_m/\tau_{NL} = \infty$, $\rho_H = 0.45$}
 \label{fig:anisotropy}
 \end{figure}

In order to quantify the level of anisotropy of these two runs we consider the 2D energy spectrum which is defined as
\begin{equation}
 E_{2D}(k_\perp,k_\parallel) = \sum_{\substack{k_\parallel \leq |\bm k \sdot \bm e_z| < k_\parallel+1 \\ k_\perp \leq |\bm k \times \bm e_z| < k_\perp+1}}|\widehat{\bm u}_{\bm k}|^2.
\end{equation}
The sum is restricted here at energy in cylinders of radius $k_\perp$ and energy in planes $k_\parallel$. %parallel to the rotation axis $z$. 
Figures \ref{fig:2Dspeca} and \ref{fig:2Dspecb} show the 2D energy spectrum for the flows with $\tau_m/\tau_{NL} = 0.5$ and $\infty$, respectively. The contours of the 2D energy spectrum for an isotropic flow is represented by concentric circles centered at the origin of the axes. Any deviation from the circular pattern indicates the level of anisotropy in the flow.
% Anisotropy
By comparing the two contour plots of $E_{2D}$, it becomes clear that in Fig. \ref{fig:2Dspecb} the energy is less concentrated close to the $k_\parallel = 0$ modes and the intermediate and small scales are closer to isotropy, implying that the flow with the time-independent forcing is overall less anisotropic than the flow with the highly random-in-time forcing. This observation is in agreement with the visualisation of Fig. \ref{fig:visualisations}, which prompt us to postulate that the helicity plays a central role on the suppression of anisotropy in the flow.

\subsection{\label{sec:specs}Spectral behaviour}
In this section we present the spectra of the energy $E(k)$ and the energy flux $\Pi_E(k)$. The energy spectrum was spherically averaged using the following expression
\begin{equation}
 E(k) = \sum_{k \leq |\bm k| < k+1} |\widehat{\bm u}_{\bm k}|^2
\end{equation}
and the spectrum of the energy flux was obtained as
\begin{equation}
 \Pi_E(k) = \sum_{k=1}^K \sum_{k \leq |\bm k| < k+1} \widehat{\bm u}^*(\bm k) \sdot \widehat{(\bm u \times \bm \omega)}(\bm k).
\end{equation}
The energy flux is a measure that illustrates the direction of the energy cascades. These spectra were time-averaged after the flows have reached a steady state solution. Note that the flow with the time-independent forcing ($\tau_m/\tau_{NL} = \infty$) was time-averaged only after 800$\tau_{NL}$ turnover times, when the new steady state was reached (see Fig. \ref{fig:injection}).

\begin{figure}
 \centering
 \begin{subfigure}{0.49\textwidth}
   \includegraphics[width=\textwidth]{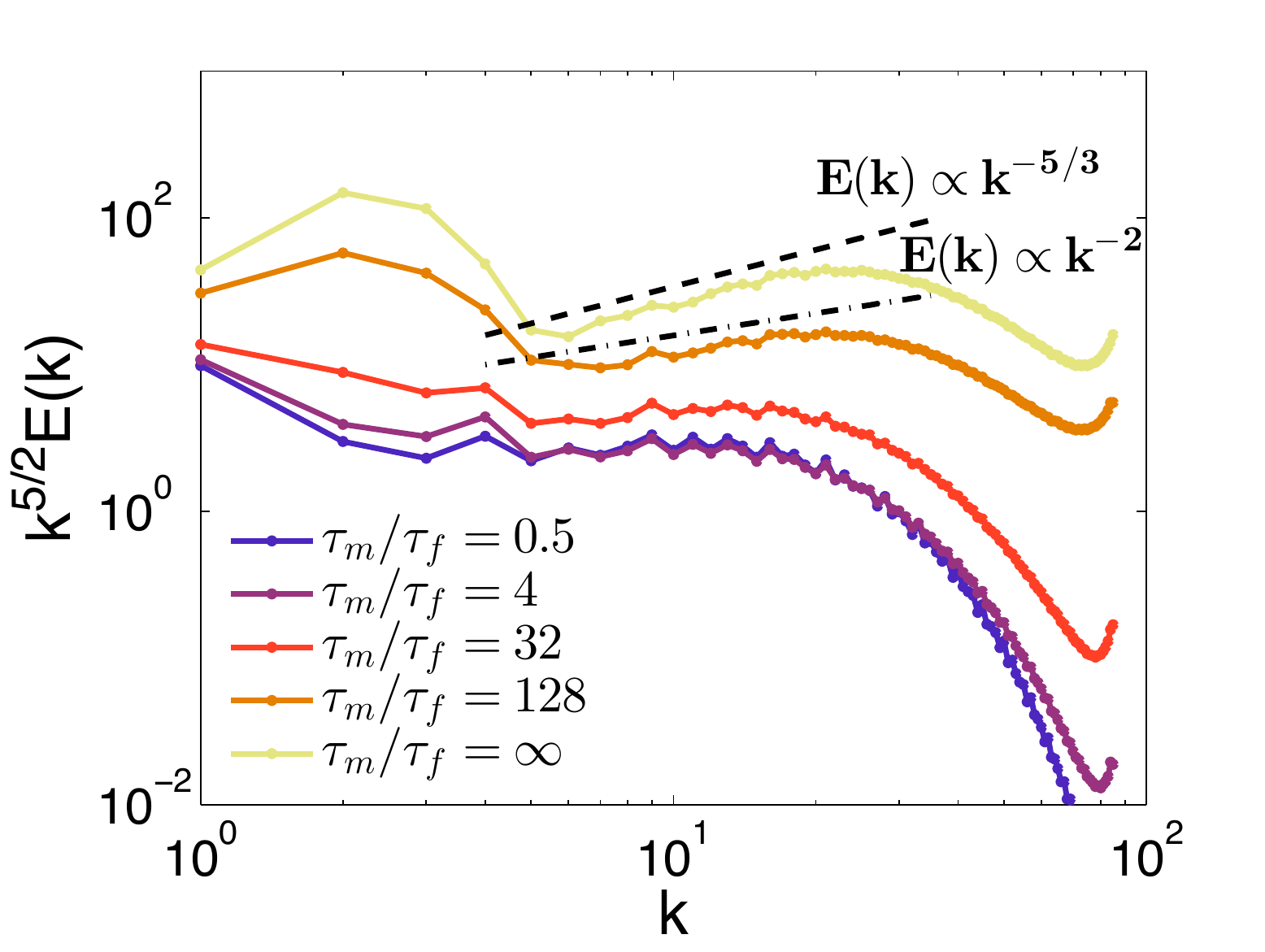}
   \caption{}
   \label{fig:Espec}
 \end{subfigure}
 \begin{subfigure}{0.49\textwidth}
   \includegraphics[width=\textwidth]{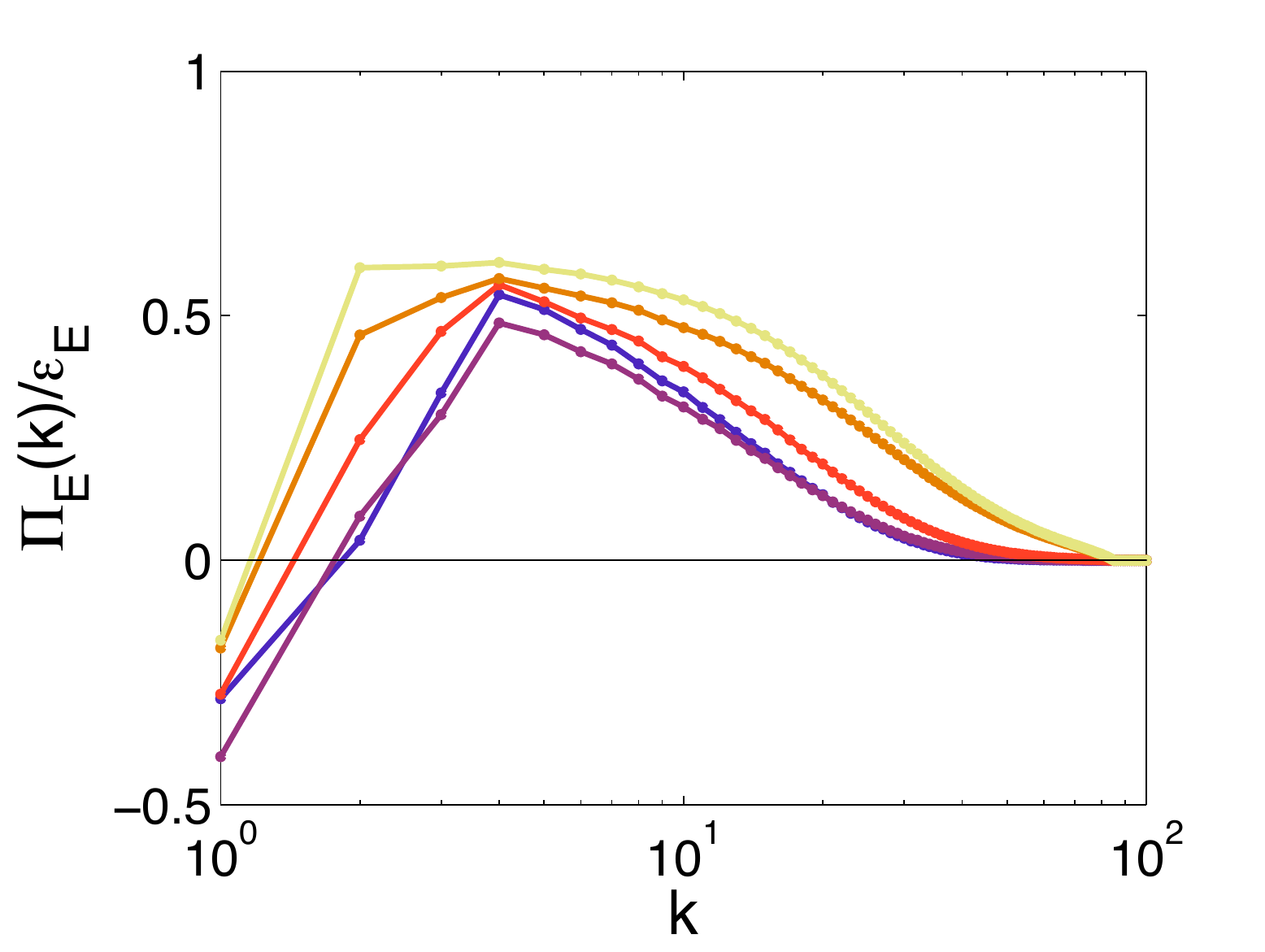}
   \caption{}
   \label{fig:fluxes}
 \end{subfigure}
  \caption{(Color online) (a) %Spherically averaged 
  Energy spectra $E(k)$ compensated by $k^{5/2}$ and (b) the energy flux spectra $\Pi_E(k)$ normalised with the dissipation rate $\epsilon_E$ for flows with different forcing memory time scale at Rossby number $Ro_f = 0.1$ and Reynolds number $Re_f = 333$}
  \label{fig:specs}
 \end{figure}

The effects of the memory time scale of the forcing is also apparent on the spectral dynamics of our flows. Figure \ref{fig:Espec} shows the energy spectra of the flows with different memory time scales of the forcing compensated by $k^{5/2}$.
The spectra of these flows obey different power laws which clearly depend on the memory time scale of the forcing. The runs that are forced with the highly random-in-time forcings have a $k^{-5/2}$ scaling. As $\tau_m$ increases the spectra start to deviate gradually from the $k^{-5/2}$ scaling towards a $k^{-2}$ and finally reach a $k^{-5/3}$ scaling for the flow with the time-independent forcing. The $k^{-5/3}$ energy spectrum can be interpreted from the fact that the intermediate and small scales of the flow are closer to isotropy (see Figs. \ref{fig:2Dspecb} and \ref{fig:vizhelb}) and hence we expect the Kolmogorov phenomenology to be valid in this case. %with the non-linear time scale $\tau_{NL}$ to dominate the dynamics responsible for the $-5/3$ scaling exponent. 
All the exponents that we observe here could be related to the various phenomenologies on strong and weak-wave turbulence in the literature, where the interplay between $\tau_{NL}$ and the time scale of the inertial waves $\tau_w \propto \Omega^{-1}$ is central to obtain the different energy spectra. These spectral exponents have also been observed in other studies of forced rotating flows \citep{yeungzhou98,mininnipouquet10,alexakis15}. So, it is clear that our results suggest a %possible 
lack of universality in forced rotating turbulence. However, higher Reynolds number computations integrated for extremely long times are necessary to verify if this is true. %Unfortunately, this is beyond our current computational capabilities.

The corresponding spectra for the energy flux $\Pi_E(k)$ normalised by the energy dissipation rate $\epsilon_E = 2\nu\sum_k k^2E(k)$ are shown in Fig. \ref{fig:fluxes}. The positive flux in this plot indicates a forward cascade while the negative flux indicates a transfer of energy from the small to the large scales of the flow. In the case of negative flux we do not talk about a cascade because we do not have enough scale separation between the forcing scale and the box size. As the memory time scale of the forcing increases, the forward cascade becomes stronger. %This is something in contradiction to what one expects in non-rotating, homogeneous and isotropic turbulence, where the forward energy cascade it is expected to slow down due to the presence of helicity \citep{kraichnan73}. 
On the other hand, the flux of energy towards the large scales increases as the forcing becomes more random-in-time with the flow reaching a quasi-2D state. These observations are in line with the visualisations of Fig. \ref{fig:visualisations} which are even at higher Reynolds numbers and thus they favour the non-universal scenario.

We already saw that, as the forcing becomes less time-dependent helicity increases considerably in our flows, so these changes in the spectra can also be related to the presence of strong net helicity in the flow. This is in agreement with prior studies that have shown the influence of helicity on the energy spectrum by directly injecting helicity into the flow \citep{mininnipouquet09,mininnipouquet10}. In contrast, helicity does not seem to have any significant effect on the spectra in non-rotating, homogeneous and isotropic helical turbulence \citep{dfa15}.

Since helicity is %a field with a non-definite sign 
not a sign-definite quantity and because we do not inject any net helicity, the sign of helicity in our flows undergoes changes in its inertial range. Therefore, there is either no power-law or difficult to define one in our helicity spectra. For this reason, we do not %give attention to the 
show any helicity spectra here. %A similar problem arises to the spectrum of cross-helicity in magnetohydrodynamic turbulence \citep{grappinetal83}.
In the next section, we examine the Rossby number dependence of the dynamics of the flows.

\section{\label{sec:Rossby}Rossby number dependence}
\subsection{\label{sec:global}Global behaviour}
In the previous sections we saw that the dynamics depend on the nature of the forcing for a given Rossby number. Here, we investigate the effects of the rotation rate on the flows, focusing on the extreme cases of the forcing being highly random-in-time ($\tau_m/\tau_{NL} = 0.5$) and time-independent ($\tau_m/\tau_{NL} = \infty$) for fixed $Re_f = 333$. Again here we restrict ourselves to moderate Reynolds numbers because extremely long integration times for the runs with time-independent forcing are inevitable. %as we have seen already.

For high Rossby number flows the effect of rotation is negligible and the energy is expected to flow to scales smaller than the forcing scale. However, as Rossby number is decreased and the flow tends to become quasi-2D, there is more and more energy transferred to scales larger than the forcing scale due to an inverse cascade \citep{pouquetetal13}.

We examine the energy and the relative helicity for runs with different Rossby numbers. The triangles and circles denote runs forced with a time-independent forcing and random-in-time forcing, respectively.
%Rossby number dependence
 \begin{figure}
 \centering
 \begin{subfigure}{0.49\textwidth}
   \includegraphics[width=\textwidth]{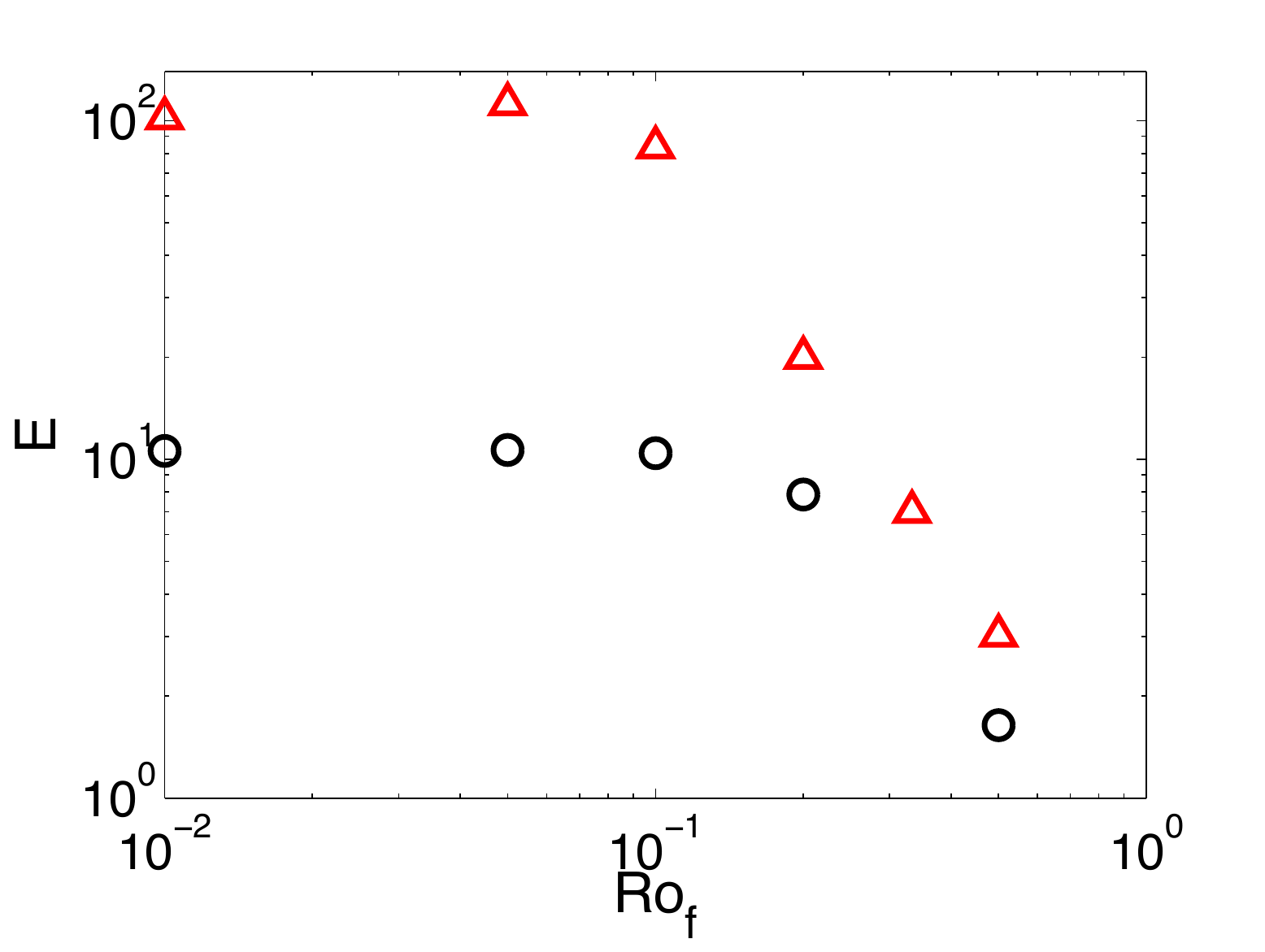}
   \caption{}
   \label{fig:Erossby}
 \end{subfigure}
 \begin{subfigure}{0.49\textwidth}
   \includegraphics[width=\textwidth]{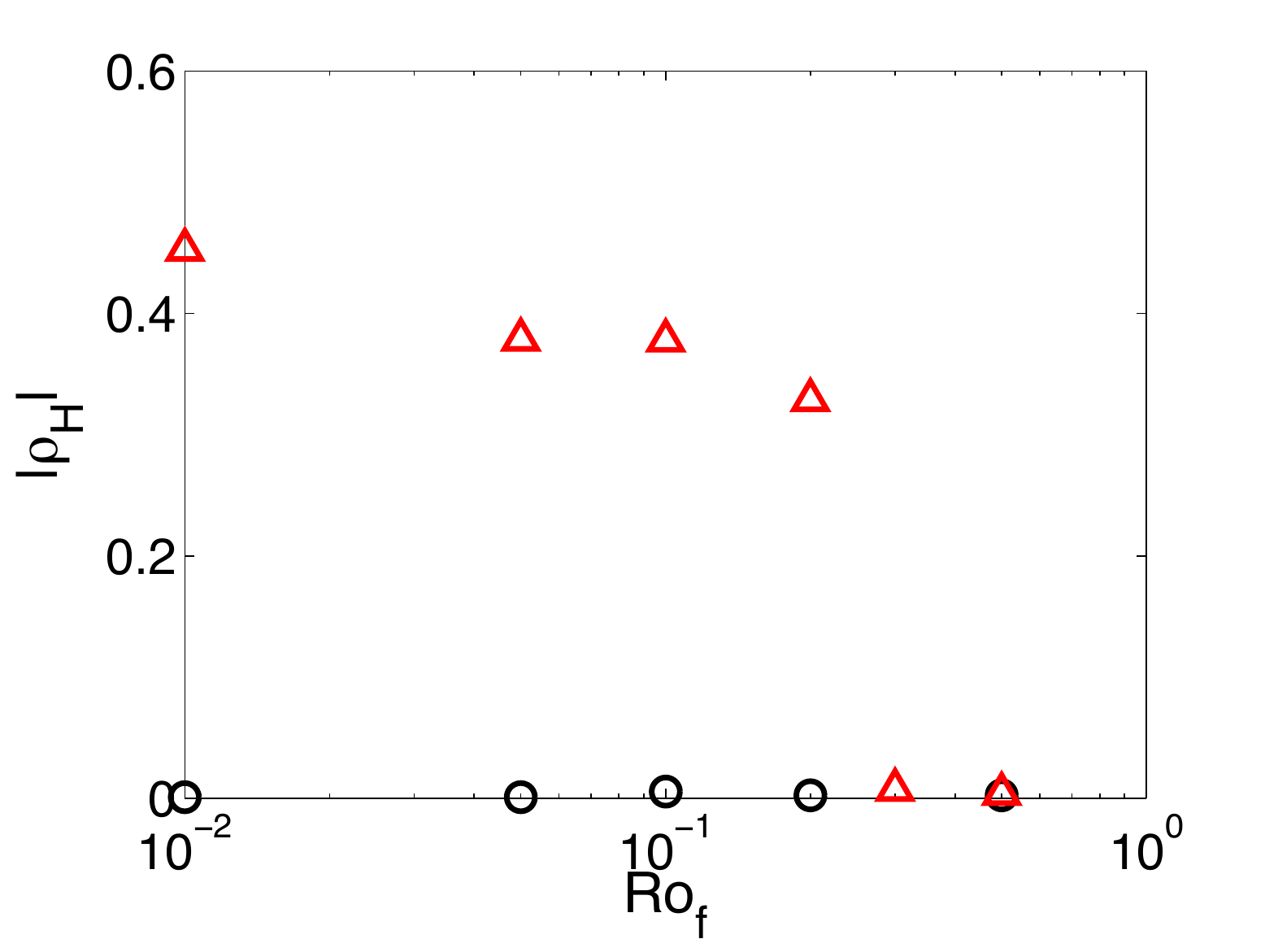}
   \caption{}
   \label{fig:Hrossby}
 \end{subfigure}
  \caption{(Color online) Rossby number dependence of (a) energy and (b) absolute value of relative helicity for flows with Reynolds number $Re_f = 333$. The $\bigcirc$ and $\triangle$ denote runs forced with $\tau_m/\tau_{NL} = 0.5$ and $\tau_m/\tau_{NL} = \infty$, respectively.}
  \label{fig:Rossby}
 \end{figure}
As Rossby number is decreased we see that energy increases as expected (see Fig. \ref{fig:Erossby}). However, the rate of increase and the values of energy for high enough rotation rates depend on the nature of the mechanical force. Note that the flow with the time-independent forcing has much more energy at small Rossby numbers.

The relative helicity behaves also very differently for the two types of flows and this is shown in Fig. \ref{fig:Hrossby}. The flow with the random-in-time forcing has zero net helicity for all $Ro_f$. However, the flow with the time-independent forcing bifurcates to a state of non-zero helicity for small enough Rossby numbers. The value of $|\rho_H|$ seems to vary discontinuously as $Ro_f$ is decreased with the flow bifurcating to a helical state at the critical $Ro_f^{crit} \simeq 0.2$. Thus, the transition from the non-helical to the helical state is a jump bifurcation. In summary, net helicity emerges in the flow only for small enough $Ro_f$ and long enough $\tau_m$.

Helicity is a pseudoscalar quantity and  $H \neq 0$ only if it is directly injected into the flow (i.e. $\bm u \sdot (\grad \times \bm f) \neq 0$) by a helical mechanical force or if another pseudoscalar quantity exists related to the pseudovector $\grad \times \bm f$. 
In our work, we observe that net helicity emerges in rapidly rotating flows with long enough memory time scale forcings only. So, a pseudovector that relates the rotation vector with the forcing is $\bm \Omega \times (\grad \times \bm f)$ and hence the pseudoscalar quantity that will allow the generation of helicity in a rotating flow is
\begin{equation}
 H \propto \bm u \sdot \bm \Omega \times (\grad \times \bm f).
 \label{eq:helicity}
\end{equation}
A similar expression was derived in a different way by \cite{hide75} for a rapidly rotating flow in geostrophic balance assuming that the non-linear term is negligible. Now, from Eq. \eqref{eq:helicity} we can deduce that no net helicity will be generated for a short memory time scale forcing since $\avg{\grad \times \bm f}_t = 0$ (with $\avg{.}_t$ denoting an average over time), assuming isotropy and ergodicity. On the other hand, for a forcing with long enough memory time scale $\avg{\grad \times \bm f}_t = \bm g(\bm x) \neq 0$ and therefore $H \neq 0$ for long enough integration time scales in agreement with our observations.

\cite{moffatt70} suggested that a random superposition of inertial waves will exhibit a lack of mirror-symmetry if and only if there is a mechanical excitation on a preferred direction in the propagation of the waves with respect to the axis of rotation. Otherwise, the random superposition of inertial waves in equal proportions would give zero net helicity. Based on Eq. \eqref{eq:helicity}, we conjecture that such a mechanism is pertinent to our flows where the angle $\varphi$ between $\Omega \bm e_z$ and $\grad \times \bm f$ is fixed at time $t=0$ for a time-independent forcing and hence such a forcing can add a preferred direction of propagation to the inertial waves inducing the mirror-symmetry breaking in our flows. From the other side, a highly random-in-time forcing can excite inertial waves on all directions in equal proportions since $\varphi$ is random in time and this is why the net helicity remains zero for any value of $Ro_f$ in this case.

\subsection{\label{sec:specs}Spectral behaviour}
\begin{figure}
 \centering
 \begin{subfigure}{0.49\textwidth}
   \includegraphics[width=\textwidth]{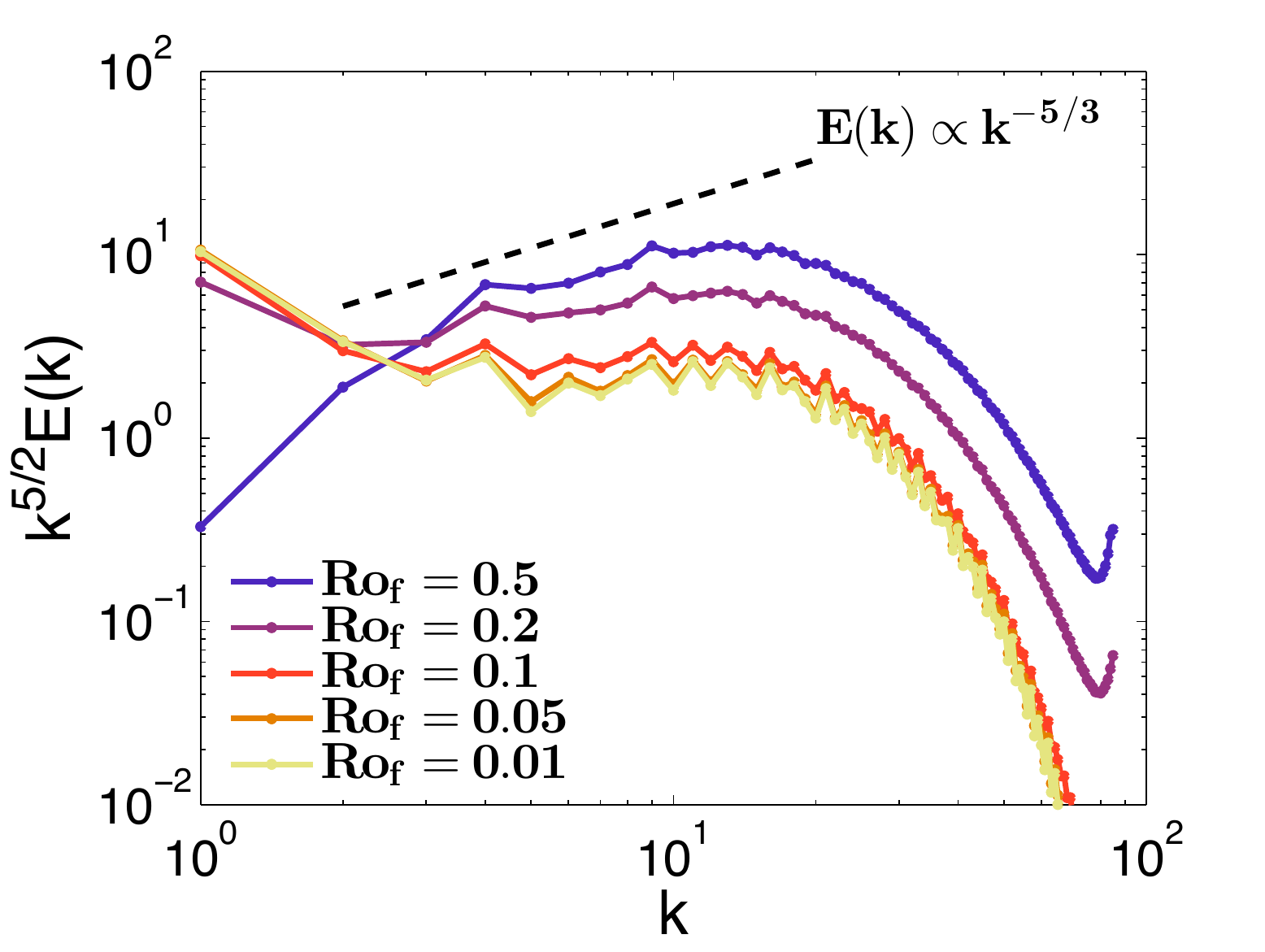}
   \caption{}
   \label{fig:Espect05}
 \end{subfigure}
 \begin{subfigure}{0.49\textwidth}
   \includegraphics[width=\textwidth]{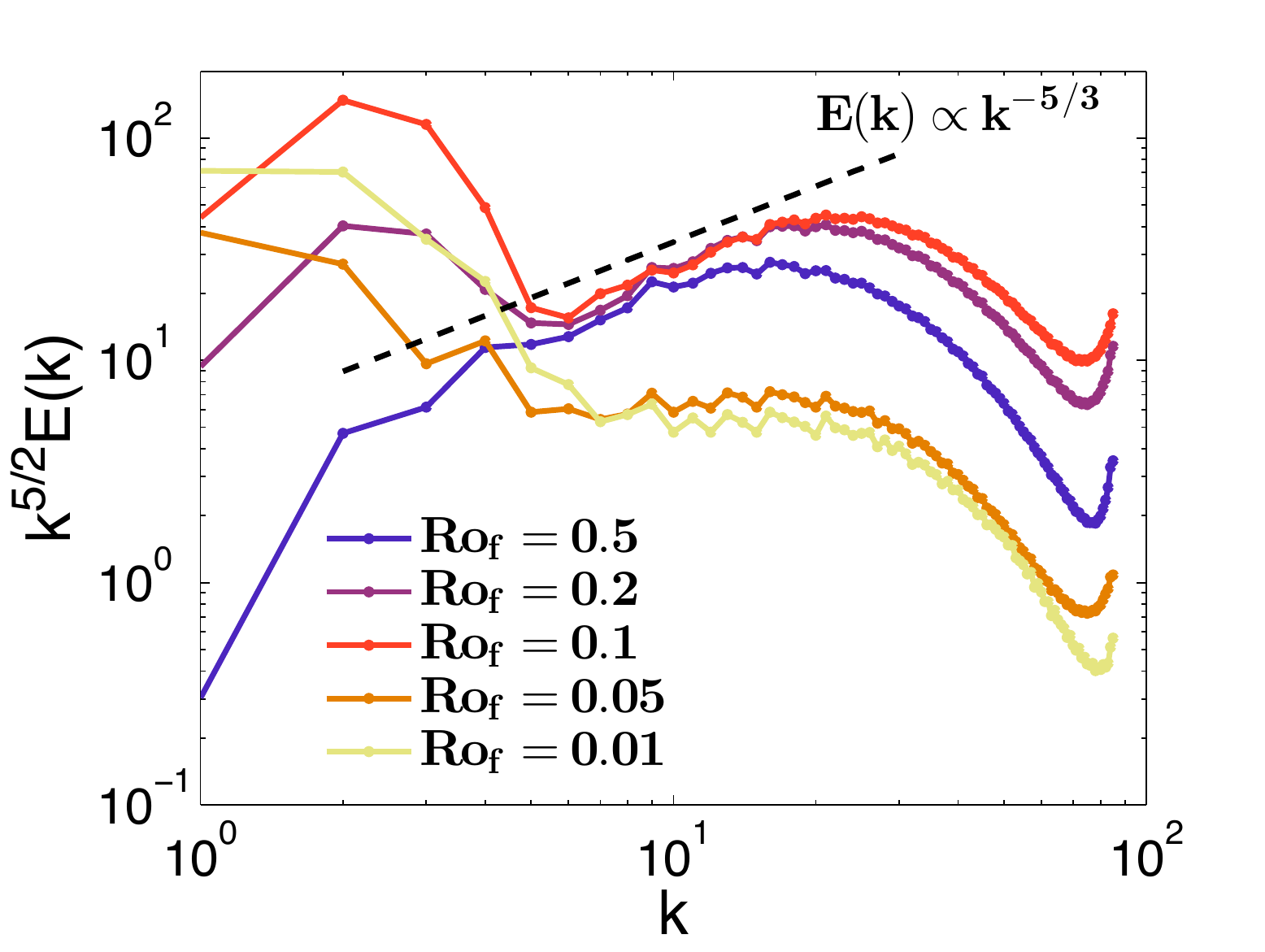}
   \caption{}
   \label{fig:EspectInf}
 \end{subfigure}
  \caption{(Color online) Energy spectra $E(k)$ compensated by $k^{5/2}$ for the flows with (a) $\tau_m/\tau_{NL} = 0.5$ and (b) $\tau_m/\tau_{NL} = \infty$ at different Rossby numbers and Reynolds number $Re_f = 333$}
  \label{fig:EspecRof}
 \end{figure}

Here we analyse the energy spectra of the flows %with highly random-in-time forcing and time-independent forcing 
at different Rossby numbers. In Fig. \ref{fig:Espect05} we present the energy spectra $E(k)$ compensated by $k^{5/2}$ of the flows with the highly random-in-time forcing. 
For $Ro_f = 0.5$ the energy spectrum is close to the Kolmogorov $k^{-5/3}$ scaling with the effects of the Coriolis force having no significant influence on the dynamics of the flow. However, as the Rossby number decreases $\tau_w \propto \Omega^{-1}$ becomes the dominant timescale and then the spectrum is changed to the weak-wave turbulence prediction of $E(k) \propto k^{-5/2}$ .

Similar behaviour is observed for the spectra of the flows with the time-independent forcing (see Fig. \ref{fig:EspectInf}) but with two different characteristics. The first is the significant condensation of energy at large scales for small enough Rossby numbers in comparison to the flows with the random-in-time forcing. The second is the transition from the Kolmogorov-like regime with $\tau_E \ll \tau_w$ to the weak-wave turbulence regime with $\tau_w \ll \tau_E$, which occurs at lower Rossby numbers, showing the dependence of this transition to the nature of the mechanical force. 

Here, we should point out that weak-wave turbulence theory arguments, which assume uniform and isotropic forcing, predict a $k^{-5/2}$ spectrum but they do not predict condensation of energy at large scales due to an inverse cascade in unbounded domains. This is in agreement with Fig. \ref{fig:Espect05} where there is some energy condensation at large scales but not significant in comparison to Fig. \ref{fig:EspectInf}. 
%However, the spectra in Fig. \ref{fig:EspectInf} that follow the $k^{-5/2}$ scaling show a considerable energy condensation at large scales. So, we do not expect necessarily weak-wave turbulence to be valid for low Rossby number flows with time-independent forcing even though $E(k) \propto k^{-5/2}$.
However, the energy condensation at larges scales of Fig. \ref{fig:EspectInf} suggests that weak-wave turbulence theory is not necessarily valid for the small Rossby number flows with time-independent forcing even though $E(k) \propto k^{-5/2}$.

\section{\label{sec:end}Discussion \& conclusions}
The dependence of the dynamics of rotating turbulence on the nature of the large scale mechanical force is studied by means of numerical simulations to shed light on the disparate results in the literature. 
For moderate Reynolds and low Rossby number flows we systematically vary the memory time scale $\tau_m$ of the mechanical force. As $\tau_m$ increases the forcing mechanism becomes less time-dependent and essentially less isotropic. %For a given Reynolds and Rossby number 
We are able to demonstrate that different steady state solutions will be reached if one is able to integrate for long enough time scales, showing the dependence of the flows on the forcing mechanism. When $\tau_m \propto \tau_{NL}$ we observe that mirror-symmetry spontaneously breaks in the flow even though our mechanical force is non-helical. %injects zero net helicity. 
Moreover, as the forcing mechanism becomes less time-dependent (long $\tau_m$) the net helicity increases. This is also true for the highest Reynolds number simulations that we carried out. We notice that helical waves break the tendency of the small and intermediate scales of the flows with the time-independent forcing to become 2D due to the imposed strong rotation. This makes the flow less anisotropic in contract to a flow with highly random-in-time forcing where the net helicity appears to be negligible. %as it is expected.

In addition, for moderate $Re_f$ and low $Ro_f$ flows %we get different 
both the power laws for the energy spectrum and the forward and inverse fluxes of energy %for the different types of 
depend strongly on the forcing mechanism. Depending on the value of $\tau_m$ we obtain $E(k) \propto k^{-5/3}$, $k^{-2}$ and $k^{-5/2}$ suggesting a lack of universality in rotating turbulence. \cite{alexakis15} showed that no matter how large the Reynolds number can be there is a small enough Rossby number such that the flow exhibits a particular behaviour (e.g. weakly rotating turbulence, quasi-2D condensates) provided that an appropriate $\alpha>0$ is considered in the scaling $Ro_f \propto Re_f^{-\alpha}$ 
%Here, we expect the exponent $\alpha$ to depend also on the mechanical force. 
(where $\alpha$ is expected to depend on the mechanical force).
So, it seems plausible %conceivable
that forced rotating turbulent flows are non-universal. To corroborate this argument a large extent of the control parameters space should be covered with higher Reynolds number simulations integrated for extremely long times. However, this is beyond %the purpose of the current study. 
the reach of current computational capabilities.

The Rossby number dependence on the dynamics of flows with a highly random-in-time and a time-independent mechanical force is also investigated at moderate Reynolds numbers. For weakly rotating turbulence (high $Ro_f$) the total energies of the two systems are comparable. Even so for small enough $Ro_f$, even though large scale vortices are present in both systems, energy condensates at large scales only for the flow with the time-independent forcing as the energy spectra demonstrate.

Moreover, for large $Ro_f$ the net helicities of the two systems are zero but as $Ro_f$ becomes smaller there is a critical $Ro_f^{crit}$ which the flow with the time-independent forcing bifurcates discontinuously from a non-helical state to a helical state. On the other hand, the helicity of the flow with the random-in-time forcing remains zero for all values of $Ro_f$. 
% \newer{Based on this result we looked at the effect of the forcing mechanism on the excitation of the inertial waves by deriving their dispersion relation taking into account the mechanical force. This derivation highlighted the importance of the angle between $\Omega\bm e_z$ and $\bm{\widehat f}$.} 
Based on this observation we argue that the angle between $\Omega\bm e_z$ and $\grad \times \bm f$ is important on the excitation of the inertial waves and concequently on the generation of net helicity in rotating flows.
Thus, a time-independent forcing adds a preferred direction of propagation to the inertial waves inducing the mirror-symmetry breaking in our flows, since this angle is fixed in time. From the other side, a highly random-in-time forcing with excites inertial waves on all directions in equal proportions and this is why the net helicity remains zero for any value of $Ro_f$. Such a mechanism has also been proposed for planetary dynamos \citep{moffatt70}.

In the end, the lack of consistency of the results in the literature is attributed here on the forcing-depend dynamics of forced rotating turbulent flows. Experiments should be able to show if this is true at higher Reynolds and lower Rossby numbers. The spontaneous emergence of helicity in such flows is an important aspect with implications on %many geophysical and astrophysical phenomena.
cyclones persistence and intensity in supercell thunderstorms, a phenomenon which defies %explanation for 
weather forecasting \citep{markowskietal98} but also in planetary dynamos.

\section*{Acknowledgements}
VD would like to thank A. Alexakis %for motivating this work and 
for enlightening discussions on rotating Taylor-Green flows.
VD would also like to thank A. Alexakis and M. Linkmann for their useful comments on the first draft of the manuscript. VD acknowledges financial support from the Newton International Fellowship funded by the Royal Society and the British Academy of Sciences. The computations were performed on ARC1 and ARC2, part of the High Performance Computing facilities at the University of Leeds, UK.
%\end{acknowledgements}

%
\bibliographystyle{jfm}
\bibliography{references}

\begin{thebibliography}{47}
\expandafter\ifx\csname natexlab\endcsname\relax\def\natexlab#1{#1}\fi

\bibitem[Alexakis(2015)]{alexakis15}
{\sc Alexakis, A.} 2015 Rotating taylor-green flow. {\em Journal of Fluid
  Mechanics\/} {\bf 769}, 46--78.

\bibitem[Andr\'e \& Lesieur(1977)]{andrelesieur77}
{\sc Andr\'e, J.~C. \& Lesieur, M.} 1977 Influence of helicity on the evolution
  of isotropic turbulence at high reynolds number. {\em Journal of Fluid
  Mechanics\/} {\bf 81}, 187--207.

\bibitem[Bartello {\em et~al.\/}(1994)Bartello, M\'etais \&
  Lesieur]{bartelloetal94}
{\sc Bartello, Peter, M\'etais, Olivier \& Lesieur, Marcel} 1994 Coherent
  structures in rotating three-dimensional turbulence. {\em Journal of Fluid
  Mechanics\/} {\bf 273}, 1--29.

\bibitem[Bewley {\em et~al.\/}(2007)Bewley, Lathrop, Maas \&
  Sreenivasan]{bewleyetal07}
{\sc Bewley, Gregory~P., Lathrop, Daniel~P., Maas, Leo R.~M. \& Sreenivasan,
  K.~R.} 2007 Inertial waves in rotating grid turbulence. {\em Physics of
  Fluids\/} {\bf 19}, 071701,.

\bibitem[Boffetta \& Ecke(2012)]{boffettaecke12}
{\sc Boffetta, Guido \& Ecke, Robert~E.} 2012 Two-dimensional turbulence. {\em
  Annual Review of Fluid Mechanics\/} {\bf 44}~(1), 427--451.

\bibitem[van Bokhoven {\em et~al.\/}(2009)van Bokhoven, Clercx, van Heijst \&
  Trieling]{vanbokhovenetal09}
{\sc van Bokhoven, L. J.~A., Clercx, H. J.~H., van Heijst, G. J.~F. \&
  Trieling, R.~R.} 2009 Experiments on rapidly rotating turbulent flows. {\em
  Physics of Fluids\/} {\bf 21}, 096601.

\bibitem[Bracco \& McWilliams(2010)]{braccowilliams10}
{\sc Bracco, Annalisa \& McWilliams, James~C.} 2010 Reynolds-number dependency
  in homogeneous, stationary two-dimensional turbulence. {\em Journal of Fluid
  Mechanics\/} {\bf 646}, 517--526.

\bibitem[Cambon {\em et~al.\/}(1997)Cambon, Mansour \& Godeferd]{cambonetal97}
{\sc Cambon, Claude, Mansour, N.~N. \& Godeferd, F.~S.} 1997 Energy transfer in
  rotating turbulence. {\em Journal of Fluid Mechanics\/} {\bf 337}, 303--332.

\bibitem[Campagne {\em et~al.\/}(2014)Campagne, Gallet, Moisy \&
  Cortet]{campagneetal14}
{\sc Campagne, Antoine, Gallet, Basile, Moisy, Frédéric \& Cortet,
  Pierre-Philippe} 2014 Direct and inverse energy cascades in a forced rotating
  turbulence experiment. {\em Physics of Fluids\/} {\bf 26}, 125112.

\bibitem[Constantin \& Majda(1988)]{constantinmajda88}
{\sc Constantin, Peter \& Majda, Andrew} 1988 The beltrami spectrum for
  incompressible fluid flows. {\em Communications in Mathematical Physics\/}
  {\bf 115}, 435--456.

\bibitem[Dallas \& Alexakis(2015)]{da15}
{\sc Dallas, V. \& Alexakis, A.} 2015 Self-organisation and non-linear dynamics
  in driven magnetohydrodynamic turbulent flows. {\em Physics of Fluids\/} {\bf
  27}, 045105.

\bibitem[Dallas {\em et~al.\/}(2015)Dallas, Fauve \& Alexakis]{dfa15}
{\sc Dallas, V., Fauve, S. \& Alexakis, A.} 2015 Statistical equilibria of
  large scales in dissipative hydrodynamic turbulence. {\em Phys. Rev. Lett.\/}
  {\bf 115}, 204501.

\bibitem[Davidson {\em et~al.\/}(2006)Davidson, Staplehurst \&
  Dalziel]{davidsonetal06}
{\sc Davidson, P.~A., Staplehurst, P.~J. \& Dalziel, S.~B.} 2006 On the
  evolution of eddies in a rapidly rotating system. {\em Journal of Fluid
  Mechanics\/} {\bf 557}, 135--144.

\bibitem[Deusebio {\em et~al.\/}(2014)Deusebio, Boffetta, Lindborg \&
  Musacchio]{deusebioetal14}
{\sc Deusebio, E., Boffetta, G., Lindborg, E. \& Musacchio, S.} 2014
  Dimensional transition in rotating turbulence. {\em Phys. Rev. E\/} {\bf 90},
  023005.

\bibitem[Galtier(2003)]{galtier03}
{\sc Galtier, S\'ebastien} 2003 Weak inertial-wave turbulence theory. {\em
  Phys. Rev. E\/} {\bf 68}, 015301.

\bibitem[G\'omez {\em et~al.\/}(2005)G\'omez, Mininni \& Dmitruk]{mpicode05}
{\sc G\'omez, Daniel~O., Mininni, Pablo~D. \& Dmitruk, Pablo} 2005 Parallel
  simulations in turbulent {MHD}. {\em Physica Scripta\/} {\bf T116}, 123--127.

\bibitem[Greenspan(1968)]{greenspan68}
{\sc Greenspan, H.~P.} 1968 {\em The Theory of Rotating Fluids\/}. Cambridge
  University Press.

\bibitem[Hide(1975)]{hide75}
{\sc Hide, R.} 1975 A note on helicity. {\em Geophysical Fluid Dynamics\/} {\bf
  7}~(1), 157--161.

\bibitem[Hopfinger \& Heijst(1993)]{hopfingervanheijst93}
{\sc Hopfinger, E~J \& Heijst, G J F~V} 1993 Vortices in rotating fluids. {\em
  Annual Review of Fluid Mechanics\/} {\bf 25}, 241--289.

\bibitem[Hossain(1994)]{hossain94}
{\sc Hossain, Murshed} 1994 Reduction in the dimensionality of turbulence due
  to a strong rotation. {\em Physics of Fluids\/} {\bf 6}, 1077--1080.

\bibitem[Kolmogorov(1941)]{k41a}
{\sc Kolmogorov, A.~N.} 1941 The local structure of turbulence in
  incompressible viscous fluid for very large {Reynolds} number. {\em Doklady
  Akademii Nauk SSSR\/} {\bf 30}, 301--305.

\bibitem[Lighthill(1965)]{lighthill65}
{\sc Lighthill, James} 1965 {\em Waves in fluids\/}. Cambridge University
  Press.

\bibitem[Maltrud \& Vallis(1991)]{maltrudvallis91}
{\sc Maltrud, M.~E. \& Vallis, G.~K.} 1991 Energy spectra and coherent
  structures in forced two-dimensional and beta-plane turbulence. {\em Journal
  of Fluid Mechanics\/} {\bf 228}, 321--342.

\bibitem[Marino {\em et~al.\/}(2013)Marino, Mininni, Rosenberg \&
  Pouquet]{marinoetal13}
{\sc Marino, Raffaele, Mininni, Pablo~D., Rosenberg, Duane \& Pouquet, Annick}
  2013 Emergence of helicity in rotating stratified turbulence. {\em Phys. Rev.
  E\/} {\bf 87}, 033016.

\bibitem[Markowski {\em et~al.\/}(1998)Markowski, Straka, Rasmussen \&
  Blanchard]{markowskietal98}
{\sc Markowski, Paul~M, Straka, Jerry~M, Rasmussen, Erik~N \& Blanchard,
  David~O} 1998 Variability of storm-relative helicity during {VORTEX}. {\em
  Monthly weather review\/} {\bf 126}~(11), 2959--2971.

\bibitem[Mininni \& Pouquet(2009)]{mininnipouquet09}
{\sc Mininni, P.~D. \& Pouquet, A.} 2009 Helicity cascades in rotating
  turbulence. {\em Phys. Rev. E\/} {\bf 79}, 026304.

\bibitem[Mininni \& Pouquet(2010)]{mininnipouquet10}
{\sc Mininni, P.~D. \& Pouquet, A.} 2010 Rotating helical turbulence. i. global
  evolution and spectral behavior. {\em Physics of Fluids\/} {\bf 22}, 035105.

\bibitem[Mininni {\em et~al.\/}(2012)Mininni, Rosenberg \&
  Pouquet]{mininnietal12}
{\sc Mininni, P.~D., Rosenberg, D. \& Pouquet, A.} 2012 Isotropization at small
  scales of rotating helically driven turbulence. {\em Journal of Fluid
  Mechanics\/} {\bf 699}, 263--279.

\bibitem[Moffatt(1970)]{moffatt70}
{\sc Moffatt, H.~K.} 1970 Dynamo action associated with random inertial waves
  in a rotating conducting fluid. {\em Journal of Fluid Mechanics\/} {\bf 44},
  705--719.

\bibitem[Moffatt(1978)]{moffatt78}
{\sc Moffatt, Henry~K.} 1978 {\em Magnetic Field Generation in Electrically
  Conducting Fluids\/}. Cambridge University Press.

\bibitem[Moisy {\em et~al.\/}(2011)Moisy, Morize, Rabaud \&
  Sommeria]{moisyetal11}
{\sc Moisy, F., Morize, C., Rabaud, M. \& Sommeria, J.} 2011 Decay laws,
  anisotropy and cyclone-anticyclone asymmetry in decaying rotating turbulence.
  {\em Journal of Fluid Mechanics\/} {\bf 666}, 5--35.

\bibitem[Morinishi {\em et~al.\/}(2001)Morinishi, Nakabayashi \&
  Ren]{morinishietal01}
{\sc Morinishi, Youhei, Nakabayashi, Koichi \& Ren, Shuiqiang} 2001 Effects of
  helicity and system rotation on decaying homogeneous turbulence. {\em JSME
  International Journal Series B Fluids and Thermal Engineering\/} {\bf 44},
  410--418.

\bibitem[Pouquet \& Mininni(2010)]{pouquetmininni10}
{\sc Pouquet, A. \& Mininni, P.~D.} 2010 The interplay between helicity and
  rotation in turbulence: implications for scaling laws and small-scale
  dynamics. {\em Philosophical Transactions of the Royal Society of London A:
  Mathematical, Physical and Engineering Sciences\/} {\bf 368}~(1916),
  1635--1662.

\bibitem[Pouquet {\em et~al.\/}(2013)Pouquet, Sen, Rosenberg, Mininni \&
  Baerenzung]{pouquetetal13}
{\sc Pouquet, A, Sen, A, Rosenberg, D, Mininni, P~D \& Baerenzung, J} 2013
  Inverse cascades in turbulence and the case of rotating flows. {\em Physica
  Scripta\/} {\bf T155}, 014032.

\bibitem[Proudman(1916)]{proudman16}
{\sc Proudman, J.} 1916 On the motion of solids in a liquid possessing
  vorticity. {\em Proceedings of the Royal Society of London A: Mathematical,
  Physical and Engineering Sciences\/} {\bf 92}~(642), 408--424.

\bibitem[Ruppert-Felsot {\em et~al.\/}(2005)Ruppert-Felsot, Praud, Sharon \&
  Swinney]{ruppertetal05}
{\sc Ruppert-Felsot, Jori~E., Praud, Olivier, Sharon, Eran \& Swinney,
  Harry~L.} 2005 Extraction of coherent structures in a rotating turbulent flow
  experiment. {\em Phys. Rev. E\/} {\bf 72}, 016311.

\bibitem[Sen {\em et~al.\/}(2012)Sen, Mininni, Rosenberg \& Pouquet]{senetal12}
{\sc Sen, Amrik, Mininni, Pablo~D., Rosenberg, Duane \& Pouquet, Annick} 2012
  Anisotropy and nonuniversality in scaling laws of the large-scale energy
  spectrum in rotating turbulence. {\em Phys. Rev. E\/} {\bf 86}, 036319.

\bibitem[Smith {\em et~al.\/}(1996)Smith, Chasnov \& Waleffe]{smithetal96}
{\sc Smith, Leslie~M., Chasnov, Jeffrey~R. \& Waleffe, Fabian} 1996 Crossover
  from two- to three-dimensional turbulence. {\em Phys. Rev. Lett.\/} {\bf 77},
  2467--2470.

\bibitem[Taylor(1917)]{taylor17}
{\sc Taylor, G.~I.} 1917 Motion of solids in fluids when the flow is not
  irrotational. {\em Proceedings of the Royal Society of London A:
  Mathematical, Physical and Engineering Sciences\/} {\bf 93}~(648), 99--113.

\bibitem[Teitelbaum \& Mininni(2011)]{teitelbaummininni11}
{\sc Teitelbaum, Tomas \& Mininni, Pablo~D.} 2011 The decay of turbulence in
  rotating flows. {\em Physics of Fluids\/} {\bf 23}~(6), 065105.

\bibitem[Tobias(2009)]{tobias09}
{\sc Tobias, S.~M.} 2009 The solar dynamo: The role of penetration, rotation
  and shear on convective dynamos. {\em Space Science Reviews\/} {\bf 144},
  77--86.

\bibitem[Tritton(1988)]{tritton88}
{\sc Tritton, D.~J.} 1988 Physical fluid dynamics. {\em Oxford, Clarendon
  Press\/} .

\bibitem[Waleffe(1992)]{waleffe92}
{\sc Waleffe, Fabian} 1992 The nature of triad interactions in homogeneous
  turbulence. {\em Physics of Fluids A\/} {\bf 4}~(2), 350--363.

\bibitem[Yarom {\em et~al.\/}(2013)Yarom, Vardi \& Sharon]{yarometal13}
{\sc Yarom, Ehud, Vardi, Yuval \& Sharon, Eran} 2013 Experimental
  quantification of inverse energy cascade in deep rotating turbulence. {\em
  Physics of Fluids\/} {\bf 25}, 085105.

\bibitem[Yeung \& Zhou(1998)]{yeungzhou98}
{\sc Yeung, P.~K. \& Zhou, Ye} 1998 Numerical study of rotating turbulence with
  external forcing. {\em Physics of Fluids\/} {\bf 10}, 2895--2909.

\bibitem[Yoshimatsu {\em et~al.\/}(2011)Yoshimatsu, Midorikawa \&
  Kaneda]{yoshimatsuetal11}
{\sc Yoshimatsu, K., Midorikawa, M. \& Kaneda, Y.} 2011 Columnar eddy formation
  in freely decaying homogeneous rotating turbulence. {\em Journal of Fluid
  Mechanics\/} {\bf 677}, 154--178.

\bibitem[Zhou(1995)]{zhou95}
{\sc Zhou, Ye} 1995 A phenomenological treatment of rotating turbulence. {\em
  Physics of Fluids\/} {\bf 7}~(8), 2092--2094.

\end{thebibliography}
\end{document}